%% file: main.tex
\documentclass[reprint,amsmath,amssymb, aps]{revtex4-2}
\usepackage{tikz}
\usepackage{braket}
\usepackage{graphicx}
\usepackage{dcolumn}
\usepackage{bm}

\usepackage{algorithm} 
\usepackage{algpseudocode}
\usepackage{amsthm}
\newtheorem{theorem}{Theorem}
\newtheorem{example}[theorem]{Example}

\newtheorem{corollary}[theorem]{Corollary}

\usepackage{xcolor}

\usepackage[normalem]{ulem}

\usepackage{tikz}
\usetikzlibrary{fit}
\usetikzlibrary{matrix}
\usepackage{tikz-3dplot}  

\usepackage[utf8]{inputenc}
\usepackage[english]{babel}
\usepackage[T1]{fontenc}
\usepackage{bm,bbold}
\usepackage{bbm}
\usepackage{amsfonts, amsmath, amsthm, amssymb} 
\usepackage{physics}
\usepackage{float}
\usepackage{hyperref}
\usepackage{color}
\usepackage{subfig}
\usepackage{comment}
\usepackage[normalem]{ulem}

\begin{document}

\preprint{APS/123-QED}

\title{Spoofing Quantum Channels Enables Low-Rank Projective Simulation}

\author{
Timothy Heightman$^{1,2}$}
\thanks{Both authors contributed equally.}
\author{Grzegorz Rajchel-Mieldzio{\'c}$^{3,1}$} \thanks{Both authors contributed equally.}

\affiliation{$^1$ICFO - Institut de Ciencies Fotoniques, The Barcelona Institute of Science and Technology, 08860 Castelldefels (Barcelona), Spain}
\affiliation{$^2$Quside Technologies SL, Carrer d’Esteve Terradas, 1, 08860 Castelldefels, Barcelona, Spain}
\affiliation{$^3$BEIT sp.\ z o.o., ul.\ Mogilska 43, 31-545 Krak{\'o}w, Poland}



\date{\today}

\begin{abstract}
The ability to characterise and discern quantum channels is a crucial aspect of noisy quantum technologies. In this work, we explore the problem of distinguishing quantum channels when limited to sub-exponential resources, framed as von Neumann (projective) measurements.
We completely characterise equivalence classes of quantum channels with different Kraus ranks that have the same outcome distributions under compatible projective measurements. In doing so, we explicitly identify gauge freedoms which can be varied without changing those compatible outcome distributions, discussing new avenues for quantum channel simulation, as well as novel adversarial strategies in noisy quantum device certification. Specifically, we show how a Sinkhorn-like algorithm enables us to find the minimum admissible Kraus rank that generates the correct outcomes. For a generic $d$-dimensional quantum system, this lowers the Kraus rank from $d^2$ to the theoretical minimum of $d$.
For up to $d = 50$, we numerically demonstrate our findings, for which the code is available and open source.  Finally, we provide an analytic algorithm for the special case of spoofing Pauli channels.
\end{abstract}

\maketitle


In the era of noisy intermediate-scale quantum (NISQ) devices, the study of how quantum states evolve in time allows us to characterise quantum hardware. Whilst closed quantum systems undergo unitary evolution, current quantum hardware contains non-unitary processes; either by design through readout, or through noise encountered in a given device. Recent attention has been given to constructing quantum hardware capable of executing unitary dynamics, with quantum error correction codes getting ever-closer to their goal of practical quantum advantage \cite{gottesman1997stabilizer, preskill1998reliable, raussendorf2012key, bell2014experimental}. 

On the other hand, much success has been found in error \textit{mitigation} in quantum hardware \cite{cai2023quantum, endo2018practical, temme2017error}. Better understanding of the true open quantum dynamics of a many-body system has facilitated improved optimal control and readout of hardware. The mathematical language to study open quantum systems is quantum channels, with the number of degrees of freedom scaling exponentially in the number of subsystems \cite{breuer2002theory}. This curse of dimensionality makes it difficult for those who wish to better understand, characterise, or simulate open quantum many-body dynamics~\cite{Terhal_1999,Narang_2007,Jung_2008,Muller-Hermes_2019,Zheng_2021,Shirokov_2020,Gaikwad_2022,Wang_2022}. Finding accurate models for such systems in known to be hard in general~\cite{Lee_2020}. 

Recent attention has thus been given to tackling this problem by introducing low-dimensional representations~\cite{Girard_2021,Girard_2022} or approximations~\cite{lancien2017approximating,Peetz_2024} of quantum channels.

In this contribution, we study the properties of gauge freedoms that arise from a restricted projective measurement set within the space of quantum channels. Importantly, these gauge freedoms give rise to equivalence classes of quantum channels that contain a variety of different Kraus ranks. As such, we can vary gauge degrees of freedom to change the Kraus rank of a channel without affecting the outcome distributions from the restricted projective measurement set. 
Our findings show and quantify the invariance of the outcome distributions, even though all the elements of the density matrix change under this transformation.

We refer to this as \textit{spoofing} a given channel with a different channel, when given access to sub-exponential resources to distinguish them. From a simulation perspective, this allows us to create a low-rank simulation of a given channel which is exact over a subset of its outcome distributions, giving a quadratic reduction in the rank and therefore, hardness of the simulability. Operationally, this opens avenues for adversarial settings whereby one can spoof a channel with one of lower Kraus rank, without being detected -- cheating users of quantum hardware out of their resource-expensive channels. 

One might argue that randomized measurement setups like classical shadows~\cite{Huang_2020,Levy_2024} can defend against this adversarial strategy. 
However, even in this setting, one may only detect the spoofers described in this contribution with some success probability. This is because whilst these measurements are \textit{technically} informationally complete (there is a non-zero probability of measuring every axis in Hilbert space), to know these distributions to a reasonable error still requires exponential resources. More concretely, we will see that a $d$-dimensional quantum channel with a randomised measurement strategy of $M$ shots would detect changes to the original $d$-dimensional channel with probability $p \sim \mathcal{O}(M/\epsilon d^2)$, up to a precision $\epsilon$. Furthermore, every architecture which can realise a randomized measurement requires a local unitary rotation which will introduce additional noise to the system. 

To summarize, if the measurement of the output state is not informationally complete, spoofing can always be exploited. In order to avoid spoofing entirely, it is therefore necessary to measure in a precise manner, requiring an exponentially large number of shots. This provides lower bounds on the necessary resources for full security when it comes to device certification.
Consequently, it is necessary to take into account these spoofing families defined by equivalence classes to prevent adversarial strategies and certify the security of quantum devices. It is also advantageous to exploit these spoofing families to find optimal strategies for low-rank simulation of quantum channels which are \textit{exact} in the outcome distributions of projective measurements.

The rest of this paper is structured as follows. We begin by recalling some basic properties of quantum channels in Section \ref{sec:quantum_channel_reps}, and formally define the quantum channel spoofing problem in Section \ref{sec:statement_of_problem}. In Section \ref{sec:parametrisation_of_spoofing}, we detail how this problem gives rise to two types of spoofing, each with different physical interpretations. 
Later in Section \ref{sec:PauliCaseStudy}, we explore an important example of how these gauge freedoms can be used to change the rank of Pauli channels, providing geometric insights to the origins of these gauge freedoms. Section \ref{sec:discussion} then offers a more extended discussion on the significance of this contribution for adversarial strategies in quantum device certification, quantum simulation, and variational quantum algorithms based on channels \cite{gravina2024adaptive}. In lieu of this discussion, we provide a numerical demonstration for finding the channel of minimal Kraus rank in a given spoofing equivalence class for a generic $d$-dimensional channel in Section \ref{sec:low_rank_sims}. We also provide an analytic algorithm for identifying the minimal Kraus rank channel in a given spoofing equivalence class for the special case of Pauli channels, formalising the examples given in Section \ref{sec:PauliCaseStudy}. Finally, we offer some concluding remarks and open problems that arise from our contribution in Section \ref{sec:conclusion}.

\subsection{Related works}
The main practical result of the present work concerns simulation of a channel with a different channel of lower Kraus rank such that the compatible outcome distributions remain identical.
This is important from the experimental perspective, as the dimension of the ancilla is equal to the Kraus rank.
Recently, Lancien and Winter~\cite{lancien2017approximating} have constructed a protocol to simulate a quantum channel with a different channel of lower rank. 
Their simulating channel is considered successful if \textit{all} outcome distributions that arise from it are $\epsilon$-close to the true distribution, using Schatten $p$-Norms as their measure of closeness. 
Whereas in Ref.~\cite{ioannou2022simulability}, the authors are able to simulate high-dimensional quantum measurements by exploiting the incompatibility of measurement operators which form an informationally complete cover of the space. In doing so, an observer would not be able to tell the difference between the simulated and true measurements since they would need access to outcome distributions corresponding to incompatible observables in order to distinguish between the two. Our work takes the philosophy behind Ref.~\cite{ioannou2022simulability}, and extends it to the more general case of quantum channels.
This provides a framework in which one can provide a simulation of a quantum channel whose accuracy is exact when looking at compatible outcome distributions, differing only in the outcomes of observables whose measurements cannot be simultaneously obtained. 

Very recently, a couple of preprints were posted exploring similar topics of channel simulation.
Refs.~\cite{Zhao_2024,Burdine_2024,Wadhwa_2024} discuss the possibility of simulating a channel with a cheaper one that is close to the original one up to some error. 
These approaches differ from our setup in a sense that spoofing is not detectable even in the limit of infinite number of measurements when restricted to a single basis.

\section{Quantum Channel Representations}
\label{sec:quantum_channel_reps}
Recall that a quantum channel 
$\mathcal{E}: B(\mathcal{H}) \rightarrow B(\mathcal{H})$ is represented as a Completely Positive Trace Preserving (CPTP) map on the space of operators, $B(\mathcal{H})$, of a Hilbert space $\mathcal{H}$. Representing the action of CPTP maps can be done in several equivalent ways. First, one may characterise a channel by its action on an input density matrix, $\rho$, as
\begin{equation}
    \mathcal{E}(\rho) = \sum_{r = 0}^R K_r \rho K_r^{\dagger},
\end{equation}
where the operators $K_r$ are the so-called Kraus operators acting on the density matrix by left- and right-matrix multiplication. Unlike unitary dynamics, these operators need only satisfy $\sum_{r = 0}^R K_r K_r^{\dagger} = \mathbb{I}$ in order for the map to be trace-preserving. We note a well-established fact that such a representation is not unique \cite{watrous2018theory}, and several sets of Kraus operators can represent the same quantum channel via the so-called unitary mixing gauge freedom. Such a transformation does \textit{not} change the action of the channel, but only its representation. 
We therefore recall the definition of the \textit{Kraus rank} as the minimum number of Kraus operators needed to represent the channel action.

An alternative representation of a channel $\mathcal{E} $ is the Choi matrix $J_\mathcal{E}$, defined as
\begin{equation}
    J_{\mathcal{E} } = (\mathbb{I}_d \otimes \mathcal{E} ) \ket{\Omega}\bra{\Omega}
    \label{eq:choi_mat_def},
\end{equation}
where $\ket{\Omega} = \frac{1}{\sqrt{d}}\sum_{i = 0}^{d - 1} \ket{i\, i}$ is the normalised, maximally entangled state in $\mathcal{H}_d \otimes \mathcal{H}_d$.

The relationship between the Choi and the Kraus representation is given via the spectral theorem. 
As a consequence, the rank of the Choi matrix is equal to the Kraus rank of the channel~\cite{watrous2018theory}. 
Since a quantum channel is a CPTP map, its Choi matrix is Hermitian and the partial trace of this Choi matrix over the second subsystem is an identity matrix of size $d$
\begin{equation}\label{eq:partial_trace}
    \mathrm{Tr}_B \, (J_{\mathcal{E}}) = \mathbb{I}.
\end{equation}\

\noindent Additionally, we recall another description of a channel via its natural (superoperator) representation. This formulates the action of a channel as a linear operation in the pseudo-vector space of quantum mixed states.
To do so, we vectorize a quantum state $\rho \rightarrow \vec{\rho}$ as $\ket{i}\!\bra{j} \mapsto \ket{i}\!\ket{j}$ and reorder the indices~\footnote{Or equivalently, vectorize row by row.}; then, the channel $\mathcal{E}$ acts as superoperator $S$ 
\begin{equation}
	S\vec{\rho} = S_{ijkl} \rho_{kl}. 
 \label{eq:naturalchannelrep}
\end{equation}
To obtain the output state $\mathcal{E}(\rho)$, one needs to ``devectorize'' the state $S\vec{\rho}$.
Operator $S$ can be thought of as a matrix of order $d^2$ or as a tensor $S_{ijkl}$, with each index taking on $d$ values.
The natural representation can be obtained from Kraus matrices $\{K_m\}_m$ through $S = \sum_m K_m \otimes K_m^*$ or from the Choi $C$ matrix via reshuffling $S = C^R$~\cite{watrous2018theory} (see App.~\ref{app:reshuffling} for more detail). 

We will use these three representations interchangeably throughout the rest of this manuscript. 
The choice of which to use is made simply for calculational ease and conceptual clarity. 
For the convenience of the reader, we summarise the relationships between the equivalent representations in Fig.~\ref{fig:channel_representation}.

\onecolumngrid
\begin{center}
\begin{figure}[H]
    \centering
    \input{figures/channel_representations}
    \caption{The equivalent representations of a quantum channel, $\mathcal{E}(\rho) = \sum_j K_j \rho K_j^{\dagger}$, where $\{K_j\}$ are the so-called Kraus operators. Here, by \textit{S.R} we refer to the Spectral Theorem \cite{watrous2018theory}, i.e., the procedure of computing the eigenvectors of the Choi matrix, and converting them to matrices via the inverse procedure of vectorisation. By $R$, we mean the reshuffling operation, explained in detail in App.~\ref{app:reshuffling}. We refer the reader to \cite{watrous2018theory} for a more detailed account of the Choi-Jamio{\l}kowski isomorphism and the representation theory that underpins this info-graphic.}
    \label{fig:channel_representation}
\end{figure}
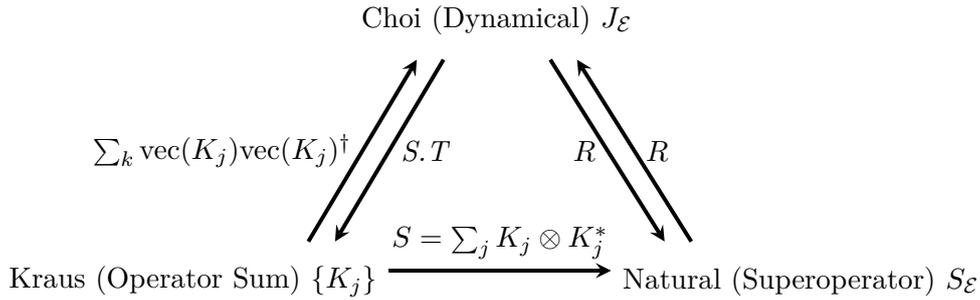
\end{center}
\twocolumngrid

Finally, we note a convenient equivalent reindexing for Choi and natural representations known as \textit{double index} notation. This is a way of indexing a $d^2 \times d^2$ matrix $M_{\mu \nu}$, where $\mu,\nu \in \{1, \ldots, d^2\}$, in terms of four indices,
\begin{equation}
    M_{\mu \nu} \rightarrow M_{ijkl},
    \label{eq:double_index}
\end{equation}
where $\mu \rightarrow i,j \in \{1,\ldots, d\},$ s.t.\ $\mu = id + b$ and $\nu \rightarrow k,l \in \{1,\ldots,d\}$ s.t.\ $\nu = kd + l$. The advantage of this re-indexing is that it takes into account the tensor-product structure of super-operators. We will also make use of this re-indexing technique throughout the manuscript.

\section{The Quantum Channel Spoofing Problem}
\label{sec:statement_of_problem}
Consider a channel acting on the space of $d$-dimensional operators, $\mathcal{E}: B(\mathcal{H}_d) \rightarrow B(\mathcal{H}_d$) with Choi matrix $J_\mathcal{E}$. We seek a secondary, different map, $\mathcal{E}': B(\mathcal{H}_D) \rightarrow B(\mathcal{H}_D)$ such that the probability of observing compatible outcomes on both channels are the same. 
The rationale for our choice stems from the fact that all measurements in a laboratory are done via projective measurements.
Those are defined via orthonormal bases; here, we focus on one of them.

We consider the scenario where this basis is the only accessible one for measurements.
However, it is possible to repeat them as many times as necessary thus obtaining the infinite, accurate statistics. The rest of the paper is devoted to study the implications that this restriction generates on the set of channels, as argued in the introduction.
We are ready to pose the central problem of this contribution.

\vspace{0.5cm}
\noindent\emph{Given a quantum channel $\mathcal{E}$ and a set of projective measurements $\mathcal{M}$, does there exist a channel $\mathcal{E}'$, whose Choi matrix $J_{\mathcal{E}} \neq J_{\mathcal{E}'}$, such that the outcome distributions of $\mathcal{M}$ remain unchanged?}
\vspace{0.3cm}

\noindent In more mathematical terms, we consider outcomes specified by POVM set \mbox{$\mathcal{M} = \{\ket{q}\!\!\bra{q}\}_ {q = 0}^{d-1}$}, in a fixed projective basis, which without loss of generality we take to be the computational basis.
We thus require that
\begin{equation}\label{eq:sameProbDiffChannel}
    p_{\mathcal{M} }(q | \mathcal{E}(\rho) ) = p_{\mathcal{M} }(q | \mathcal{E}'(\rho) ) , \;\; \forall \ket{q}\!\!\bra{q} \in \mathcal{M}, \; \rho \in B(\mathcal{H}),
\end{equation}
which defines an equivalence relation
\begin{equation}\label{eq:equivalence_relation}
    \mathcal{E} \sim_{\scalebox{0.5}{$\mathcal{M}$}} \mathcal{E}',
\end{equation}
between channels $\mathcal{E}$ and $\mathcal{E}'$.
Note here that since the measurement set is not informationally complete, not all characteristics of an output state can be gathered.
In particular, by measuring in a fixed basis we obtain diagonal elements of the output state in the basis of the projector. 

Alternatively, instead of considering the indistinguishability of all the measurements in a single basis, we could focus on the description of the evolution of the observables.
This is done via means of the adjoint map~\footnote{Notably, the adjoint map is not necessarily a quantum channel, as it can map states to non-physical states.} $\mathcal{E}^\dagger$ to the original channel $\mathcal{E}$, which is defined as the unique linear map that satisfies
\begin{equation}
    \langle \mathcal{E}^\dagger(\rho),\sigma \rangle = \langle \rho,\mathcal{E}(\sigma) \rangle,
\end{equation}
for all density matrices $\rho$ and $\sigma$.  
For us, however, the adjoint map is interesting because it allows for the evolution of the observables rather than the states. 

We consider two channels $\mathcal{E}  \neq \mathcal{E}' $, to be different when their Choi matrices,
\begin{equation}
    J_{\mathcal{E} } \neq J_{\mathcal{E}' }
    \label{eq:diffChoi},
\end{equation}
are not the same, as several Kraus representations may have the same channel action \cite{watrous2018theory}. 
To summarise, we seek to classify the sets of quantum channels satisfying $p_{\mathcal{M} }(q | \mathcal{E}(\rho) )  = p_{\mathcal{M} }(q | \mathcal{E}'(\rho) )$ for any input state $\rho$ with computational basis measurements.

For the rest of the paper, we assume that $\mathcal{M}$ is defined through the computational basis.
We provide an explicit statement of the main problem in terms of Kraus representation in App.~\ref{app:conditions_in_kraus}.

\section{Parametrization of Spoofing Channels}\label{sec:parametrisation_of_spoofing}
In this section, we explain how the problem defined in Sec.~\ref{sec:statement_of_problem} gives rise to two distinct classes of spoofer shown in Fig.~\ref{fig:spoofers}, hereby referred to as Type-I and Type-II. 

\onecolumngrid
\begin{center}
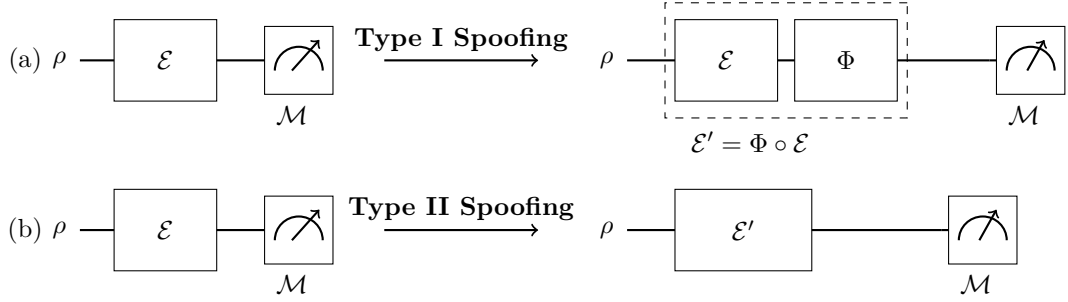
\begin{figure}[H]
    \centering
    \input{figures/spoofing_types.tex}
    \caption{Two inequivalent types of spoofing referred to as Type-I Spoofing (a), and Type-II Spoofing (b). In Type-I Spoofing, we assume we have access to the output of a given channel $\mathcal{E}$ and are able to concatenate a spoofing channel $\Phi$ to it, such that the resultant channel, $\mathcal{E}' = \Phi \circ \mathcal{E}$ is divisible into these two constituent parts. The resultant channel in (a) is depicted inside the dashed line. In Type-II Spoofing, we assume no such access to $\mathcal{E}$. Here, the spoofing channel is not necessarily divisible.}
    \label{fig:spoofers}
\end{figure}
\end{center}
\twocolumngrid

The key intuition behind there being two classes is that quantum channels are not always reversible. 
So, if we have a pair of different channels with natural representations $S_\mathcal{E} \neq S_\mathcal{E}'$, we cannot guarantee that the matrix $M$ defined via 
\begin{equation}
    S_{\mathcal{E}'} = M \cdot S_{\mathcal{E}},
\end{equation}
defines a valid channel~\footnote{To elaborate, it would require that $M = S_{\mathcal{E}'} \cdot S_{\mathcal{E}}^{-1}$ is a natural representation of some channel. 
This is not the case in general for non-unitary processes, where $\mathcal{E}^{-1}$ is not necessarily a valid channel.}.
Formally, the structure and distinction between Type-I and Type-II spoofers comes from the fact that quantum dynamics is not strictly bi-divisible \cite{Wolf_2008,davalos2023quantum}.

In Type-I spoofing, we consider the scenario where we have a (possibly unknown) channel $\mathcal{E}$, whose outcome distributions of projective measurements can equivalently be obtained from a divisible channel $\mathcal{E}' = \Phi \circ \mathcal{E}$.
On the other hand, Type-II spoofing seeks to replace a given (and known) channel, $\mathcal{E}$, with a new channel $\mathcal{E}'$, which is not necessarily divisible. While this difference may be subtle, we see a profound difference in the number gauge freedoms that each type allows, as well as their use-cases in adversarial settings, efficient quantum channel simulation, and variational quantum algorithms. See Sec.~\ref{sec:discussion} for further discussion.

We now seek to understand the structure of the spoofing channels for Type-I and Type-II by characterizing their gauge degrees of freedom. 
Here, by a gauge we mean the freedom to vary the value of parameters of the channel without changing the outcome distribution in line with Sec.~\ref{sec:statement_of_problem}.

\subsection{Type-I Spoofing}
\label{sec:TypeISpoofing}
In accordance with Fig.~\ref{fig:spoofers}, let the original channel be denoted $\mathcal{E}$, and the spoofing channel, $\mathcal{E}'$, be the result of concatenating $\mathcal{E}$ with some other channel $\Phi$. That is, 
\begin{equation}
    \mathcal{E}' = \Phi \circ \mathcal{E}.
    \label{eq:TypeISpoofer}
\end{equation}
The natural representation, $S_{\Phi}$ must act as the identity channel,  $S_{\mathbb{I}}$, over all measurable components of $\rho$. As such, the gauges in $S_{\Phi}$ are its diagonal elements, each with modulus smaller than one. 

\begin{theorem}\label{thm:gauge_freedoms}
    The natural representation of the Type-I spoofing channel, $S_{\Phi}$, is a diagonal matrix. 
\end{theorem}

\begin{proof}
In the basis of the measurement set, we require the diagonal of the output density matrix to be the same when applying $\mathcal{E}$ or $\mathcal{E}'$ in order to generate the same outcome distributions. Thus, $S_{\Phi}$ must not cause changes to $\mathcal{E}$ on entries which will affect observable components of the density matrix. 
When applied to a state, $\rho$, in vectorised form (as per the left-hand side of Eq.~(\ref{eq:naturalchannelrep})), this constrains some rows of $S_{\Omega}$.
In particular, the rows that act on observable parts of $\rho$ must be the same as the identity matrix. On other rows however, 
$S_{\Omega}$ is free to vary, implying the following block structure,
\tikzset{highlight/.style={rectangle,
                           draw=blue,
                           rounded corners = 0.5 mm, 
                           inner sep=1pt,
                           fit=#1}
}
\begin{equation}
S_{\Phi} = 
\begin{tikzpicture}[baseline=0pt]
  \matrix (m)[
    matrix of math nodes,
    nodes in empty cells,
    minimum width=width("88"),
    left delimiter=(,
    right delimiter=),
  ] {
    1 & ... & 0 & ... & 0 \\
    & & & &  \\
    & & & & \\
    0 & ... & 1 & ... & 0 \\
    & &  & & \\
    & &  & &  \\
    & & & & \\
    0 & ... & 0 & ... & 1 \\
  } ;
  \draw (m-3-1.south west) rectangle (m-2-5.north east);
  \draw (m-7-1.south west) rectangle (m-5-5.north east);
  \node (a) at (0,0.56) {\scalebox{0.8}{$ (d+1)\times d^2$}};
  \node (a) at (0,-0.4) {\scalebox{0.7}{$ \ddots $}};
\end{tikzpicture},
\end{equation}
where the rectangular blocks correspond to free sub-matrices of size $(d + 1) \times d^2$. 
The form of each block stems from the way in which a density matrix is vectorised; these blocks act on the components of $\vec{\rho}$ which correspond to off-diagonal terms in the matrix representation decomposed in the basis of our compatible measurements.

We also require that $\Phi$ is a CPTP map, adding constraints for trace preservation, and a Hermitian output. These conditions are simpler to establish by switching to the Choi representation, since the channel is CPTP if and only if its Choi matrix is positive semi-definite \cite{watrous2018theory}. 
In the Choi representation, we may enforce this via principal minors.
In particular, Sylvester’s criterion states that a Hermitian matrix is positive semidefinite if and only if all its principal minors are nonnegative. 
Thus, applying Sylvester's criterion to $J_{\Phi} = (S_{\Phi})^R$, we find that the only remaining free parameters are those that lie on the diagonal in the natural representation, $S_{ab}$. 
That is, the set of elements $\{S_{22},\ldots,S_{d^2-1,d^2-1}\}$ corresponds to gauge freedoms of the channel in the natural representation
\begin{equation}\label{eq:type_I_degrees_freedom}
    S_{\Phi} = \text{diag}(1, a_1, ... , a_d, 1, a_{d+1}, ..., a_{d^2-d},1).
\end{equation}
\end{proof}
\noindent In order to enforce trace preservation, the above gauge degrees of freedom must be bounded with modulus less than one, $|a_i| \leq 1$, while the Hermiticity of the Choi matrix provides the last condition on certain elements $a_i = a^*_j$. This leads to the following corollary.

\begin{corollary}\label{cor:number_degrees_freedom_type-I}
    The number of free real parameters of Type-I spoofing equivalence class of a channel on $d$-dimensional states is $d^2 - d$.
\end{corollary}
\begin{proof}
    There are $d^2 - d$ complex parameters that stem from Eq.~(\ref{eq:type_I_degrees_freedom}). 
    The trace-preserving condition does not lower the dimensionality of the equivalence class but only restricts its accessible values.
    However, the Hermiticity constraint on the Choi matrix adds $d^2 - d$ conditions on complex parameters $a_i$s, which concludes the proof.
\end{proof}
\noindent We end the subsection with a characterisation of the Type-I spoofing equivalence family for 1-qubit channels.

\begin{example}
    In Type-I spoofing, the gauge freedoms for an arbitrary 1-qubit channel consist of a single complex parameter, $a \in \mathbb{C}$, in the Choi representation of channel $\Phi$ from Eq.~(\ref{eq:TypeISpoofer}) (see also Fig.~\ref{fig:spoofers})
\begin{equation}
J_{\Phi} = 
\begin{pmatrix}
        1 & 0 & 0 & a \\
   0 & 0 &0& 0 \\
   0 & 0 & 0 & 0\\
   a^*  & 0 & 0 & 1 \\
\end{pmatrix}.
\end{equation}
\end{example}

\subsection{Type-II Spoofing}
\label{sec:TypeIISpoofing}

Consider the scenario in which the simulation of $\mathcal{E}$ is done by an entirely different channel.
We therefore seek a new channel $\mathcal{E}'$ such that its action alone on any input state produces the same outcome distributions as $\mathcal{E}$ with respect to the compatible measurement set $\mathcal{M}$, i.e., satisfying Eq.~(\ref{eq:sameProbDiffChannel}).
We note that Type-II spoofing needs prior knowledge of the given channel, but can be constructed for any given input channel.

\begin{theorem}\label{thm:type_II_spoofing}
    Consider the Type-II spoofing equivalence class of a $d$-dimensional channel $\mathcal{E}$ (Fig.~\ref{fig:spoofers}). 
    Then, a channel $\mathcal{E}'$ belongs to the equivalence class defined by Eqs.~(\ref{eq:sameProbDiffChannel}) and (\ref{eq:equivalence_relation}) if and only if the diagonal blocks of its Choi matrix $J_{\mathcal{E}'}$, each of size $d\times d$, are equal to the diagonal blocks of $J_{\mathcal{E}}$
\end{theorem}

\begin{proof}
    We begin with an observation concerning the natural representation: if $\mathcal{E}'$ belongs to the same spoofing equivalence class as the original channel $\mathcal{E}$, then for every input state $\rho$, the output states for $\mathcal{E}$ and $\mathcal{E}'$ must agree in the diagonal elements,
    \begin{equation}
        \forall\, q: \,\,\ \bra{q}\mathcal{E}(\rho)\ket{q} = \bra{q}\mathcal{E}'(\rho)\ket{q}.
    \end{equation}
    This means that in the natural (superoperator) representation, the two vectors $S_\mathcal{E} \, \vec{\rho}$ and $S_{\mathcal{E}'} \,\! \vec{\rho}$, agree on the elements which correspond to diagonal elements of output density matrices. 
    Using double index notation from Eq.~(\ref{eq:double_index}), this means that when $i=j$,
    \begin{equation}\label{eq:diagonal_indices}
        [S_\mathcal{E}\, \vec{\rho}\,]_{ij,kl} = [S_{\mathcal{E}'}\,\! \vec{\rho}\,]_{ij,kl} \,\,\ \forall\, k,l.
    \end{equation}
    Since this must be true for all input states, $\rho$, the equality must hold at the level of $S_\mathcal{E}$ on the appropriate rows. That is,
    \begin{equation}
         \forall\, k,l: \,\,\ [S_\mathcal{E}]_{ij,kl} = [S_{\mathcal{E}'}]_{ij,kl},
    \end{equation}
    when $i = j$.
    To obtain the Choi matrix of both channels, we may apply reshuffling (see Fig.~\ref{fig:channel_representation} and App.~\ref{app:reshuffling}), yielding
    \begin{equation}
        \forall\, j,l: \,\,\ [J_\mathcal{E}]_{ij,kl} = [J_{\mathcal{E}'}]_{ij,kl}
    \end{equation}
    when $i = k$, which is exactly the condition for equality of the diagonal blocks. 
    Observe that whenever $i \neq k$, the only requirements for blocks stem from demanding that $J_{\mathcal{E}'}$ be a valid channel.
    
    For the converse direction, we must prove that if all the diagonal blocks of the Choi matrices are equal, then the channels belong to the same Type-II equivalence class, \mbox{$\mathcal{E} \sim_{\scalebox{0.5}{$\mathcal{M}$}} \mathcal{E}'$}.
    The equality of diagonal blocks of the Choi matrices lead to the equality of rows of the natural representations that correspond to Eq.~(\ref{eq:diagonal_indices}).
    From this follows that for any input state the diagonal part of the output density matrix is equal independently on the used channel, $\mathcal{E}$ or $\mathcal{E}'$.
    This proves the equivalence in both directions.
\end{proof}

\noindent In the matrix language, condition (1) of Thm.~\ref{thm:type_II_spoofing} means that the Choi matrix of $\mathcal{E}'$ is of the form
\begin{equation}
J_{\mathcal{E}'} = 
\begin{tikzpicture}[baseline=0pt]
  \matrix (m)[
    matrix of math nodes,
    nodes in empty cells,
    minimum width=width("88"),
    left delimiter=(,
    right delimiter=),
  ] {
        &  &\circ & \circ &\circ & \circ & \dots \\
    & &\circ & \circ&\circ & \circ  & \dots\\
   \circ & \circ &  & &\circ& \circ & \dots\\
   \circ & \circ &  & &\circ& \circ & \dots\\
   \circ & \circ & \circ & \circ & & & \dots\\
   \circ & \circ & \circ & \circ & & & \dots\\
   \vdots & \vdots & \vdots & \vdots & \vdots& \vdots & \ddots\\
  } ;
  \draw (m-2-1.south west) rectangle (m-1-2.north east);
  \draw (m-4-3.south west) rectangle (m-3-4.north east);
  \draw (m-6-5.south west) rectangle (m-5-6.north east);
  \node (a) at (-1.14,1.09) {\scalebox{0.7}{$ d\times d$}};
  \node (a) at (-0.34,0.3) {\scalebox{0.7}{$ d\times d$}};
  \node (a) at (0.48,-0.48) {\scalebox{0.7}{$ d\times d$}};
\end{tikzpicture},
\end{equation}
where the boxes are of size $d\times d$ and are constrained to take the same values as the Choi matrix of the original channel, $J_{\mathcal{E}'}$. It is for this reason that $\mathcal{E}$ must be known by a Type-II spoofer; the $d\times d$ are explicitly dependent on $J_\mathcal{E}$. The rest of the elements, denoted by dots $\circ$, are free to vary provided they satisfy conditions (2) and (3). 
This leads to the following corollary.

\begin{corollary}
\label{cor:dof_type_ii}
    The number of free real parameters of the Type-II spoofing equivalence class of a channel on $d$-dimensional states is $d^4 - d^3 - d^2 + d$.
\end{corollary}

\begin{proof}
    The proof is based on the counting of the constraints. 
    First, observe that a $d$-dimensional quantum channel admits a Choi matrix representation of \emph{a priori} $d^4$ complex elements. 
    The Hermiticity of the Choi matrix leaves $d^4$ real parameters. Next, its partial trace must satisfy, Eq.~(\ref{eq:partial_trace}), introducing $d^2$ constraints. 
    
    We then introduce the constraints arising from the $d \times d$ diagonal blocks which must remain the same in both $J_{\mathcal{E}}$ and $J_{\mathcal{E}'}$. Since there are $d$ copies of these blocks, we would naively assume that this introduces a further $d^3$ constraints. However, we would then be over-counting the Hermiticity constraint, as these $d$ sub-blocks are themselves Hermitian. Therefore, keeping these blocks identical and Hermitian presents $d^3 - d$ constraints. 
    Combining the above, we see that the number of free real parameters is $d^4 - d^3 - d^2 + d$.
    We know these parameters are linearly independent by virtue of their action on the Choi matrix basis.
    
\end{proof}
\begin{example}
\label{eg:type-II_one_qubit}
    In the simplest example of one-qubit case, the spoofing freedom for a channel consists of 3 complex (6 real) parameters in the Choi representation
\begin{equation}
\begin{tikzpicture}[baseline=0pt]
  \matrix (m)[
    matrix of math nodes,
    nodes in empty cells,
    minimum width=width("88"),
    left delimiter=(,
    right delimiter=),
  ] {
     &  & a & b  \\
    & & c & -a \\
   a^* & c^* &  & \\
   b^*  & -a^* &  &  \\
  } ;
  \draw (m-2-1.south west) rectangle (m-1-2.north east);
  \draw (m-4-3.south west) rectangle (m-3-4.north east);
\end{tikzpicture},
\end{equation}
where two empty boxes contain the elements of $2\times 2$ diagonal blocks, whose values are determined by the original Choi matrix $J_\mathcal{E}$.
\end{example}
\noindent Lastly, let us remark that using the adjoint map allows for an alternative description of the spoofing freedom. 
Namely, for all the channels, $\mathcal{E}'$, in a given Type-II spoofing equivalence class, the adjoint maps $\mathcal{E}'^\dagger$ will agree on the computational basis states, i.e.,\
\begin{equation}
    \mathcal{E} \sim_{\scalebox{0.5}{$\mathcal{M}$}} \mathcal{E}'\;\; \Longleftrightarrow  \;\; \forall\, i: \;\; \mathcal{E}^\dagger(\ket{i}\!\!\bra{i}) = \mathcal{E}'^\dagger(\ket{i}\!\!\bra{i}).
\end{equation}
The above conditions means that the diagonal blocks of the Choi matrices of the adjoint maps also agree, in strict analogy to condition (1) from Theorem.~\ref{thm:type_II_spoofing}.

\section{Case study: Pauli Channels and the Computational Basis}\label{sec:PauliCaseStudy}
In this section, we show how to identify the gauge freedoms from Sec.~\ref{sec:parametrisation_of_spoofing} for Pauli channels, defined in the Kraus representation as,
\begin{equation}
    \mathcal{E}(\rho) = \sum_j \alpha_j P_j \rho P^{\dagger}_j,
\end{equation}
where $P_j \in \mathcal{P}_N$ is a member of the Pauli group on $N$ qubits. We choose the computational basis as our example of the projective measurement set as an instructive example. This is because any implemention of IC-POVMS on Pauli channels requires prohibitive resources.
The number of measurement shots required to yield an estimator, $\hat{J}_{\mathcal{E}}$ for the Choi matrix $J_{\mathcal{E}}$  such that any valid distance with respect to the true channel satisfies,
\begin{equation}
    ||J_{\mathcal{E}} -\hat{J}_{\mathcal{E}} || \leq \epsilon,
\end{equation}
scales as $\mathcal{O}(d^4/\epsilon^2)$, where $d$ is the dimension of the system's Hilbert space, and $\epsilon \in \mathbb{R}^+$. This scaling comes from the Choi matrix having $\mathcal{O}(d^4)$ independent parameters (for a channel acting on a $d$-dimensional system). To estimate the independent parameters, the number of measurements needed typically scales inversely with the square of the desired accuracy as per the standard quantum limit \cite{braginsky1995quantum}. Even with more modern (theoretically optimal) techniques such as Heisenberg-limited measurements \cite{zwierz2012ultimate, demkowicz2012elusive}, this quartic overhead in the system dimension persists.  

For channels comprising noisy qubit systems of $N$ qubits (i.e., mixed unitary channels), this scaling makes characterising a quantum channel exponentially prohibitive, since $d$ scales as $\mathcal{O}(2^N)$, hence the \texttt{NP}-hardness in distinguishing mixed unitary channels \cite{Lee_2020}. Here, we motivate the simple example of using the computational basis on a small number of qubits as an example of how the gauge freedoms arise when one does not measure in an informationally complete manner.
This is a reasonable assumption for $N$-qubit Pauli channels, for which number of elements of IC POVMs scales as $\mathcal{O}(4^N)$. The results of this section, and for the general case of $d$-dimensional channels can be found in Table~\ref{tab:free_parameters}. Further discussion on the number of shots required to detect a spoofer up to some probability, $p$, is reserved for Sec.~\ref{sec:discussion}.

\begin{center}
\begin{table}[h!] 
\centering
  \begin{tabular}{|c |c |c |c |c |}
    \hline
    & $N$ qubits   & qu$d$it \\ \hline
    Type-I Pauli& $(4^N - 2^N)/2$   & ---  \\ \hline
    Type-I general& $4^N - 2^N$   & $d^2-d$ \\ \hline
    Type-II Pauli& $4^N - 2^N$   & --- \\ \hline
    Type-II general & $16^N- 8^N-4^N+2^N$   & $d^4-d^3-d^2+d$ \\ \hline
  \end{tabular}
  \caption{Number of free real parameters for $N$ qubits and various types of spoofing, as well as for $d$-dimensional systems (qudits). 
  Note that an $N$-qubit state can always be represented as a single $2^N$-dimensional system.}
\label{tab:free_parameters}
\end{table}
\end{center}

\subsection{One Qubit}
\noindent Consider the channel $\mathcal{E}: \mathcal{B}(\mathcal{H}_2) \rightarrow \mathcal{B}(\mathcal{H}_2)$,
\begin{equation}
    \mathcal{E}(\rho) = \alpha_0 \rho + \alpha_1 X \rho X + \alpha_2 Y \rho Y + \alpha_3 Z \rho Z, 
\end{equation}
where the coefficients, $\alpha_j$ sum to one, $\rho \in \mathcal{B}_+(\mathcal{H}_2)$ is a input state, and $\{X, Y, Z\}$ are the Pauli matrices.
Using Eq.~(\ref{eq:choi_mat_def}), the Choi matrix of the channel reads
\begin{equation}
    J_{\mathcal{E}} = 
    \begin{pmatrix}
        \alpha_0 + \alpha_3 & 0 & 0 & \alpha_0 - \alpha_3 \\
        0 & \alpha_1 + \alpha_2 & \alpha_1 - \alpha_2 & 0 \\
        0 & \alpha_1 - \alpha_2 & \alpha_1 + \alpha_2 & 0 \\
        \alpha_0 - \alpha_3 & 0 & 0 & \alpha_0 + \alpha_3
    \end{pmatrix}.
\end{equation}
We may now explore our two types of spoofer on this one qubit channel.
\subsubsection*{Type-I Spoofing}
\noindent Theorem~\ref{thm:gauge_freedoms} forces the structure of the component $\Phi$ of spoofing channel, $\mathcal{E}'$, in the natural representation to be
\begin{equation}
    S_\Phi = 
    \begin{pmatrix}
        1 & 0 & 0 & 0  \\
        0 & \beta & 0 & 0 \\
        0 & 0 & \beta & 0 \\
        0 & 0 & 0 & 1
    \end{pmatrix},
\end{equation}
such that
\begin{equation}
    \begin{split}
    &J_{\mathcal{E}'} = J_{\Phi \circ \mathcal{E}} \\
    & = \begin{pmatrix}
        \alpha_0 + \alpha_3 & 0 & 0 & \beta(\alpha_0 - \alpha_3) \\
        0 & \alpha_1 + \alpha_1 & \beta(\alpha_1 - \alpha_2) & 0 \\
        0 & \beta(\alpha_1 - \alpha_2) & \alpha_1 + \alpha_2 & 0 \\
        \beta(\alpha_0 - \alpha_3) & 0 & 0 & \alpha_0 + \alpha_3
    \end{pmatrix}.
\end{split}
\end{equation}
Notice that this induces the following change in the coefficients, $\{\alpha_j\}$, of the channel,
\begin{equation}
\begin{split}
    \alpha_1 - \alpha_2 &\rightarrow \beta(\alpha_1 - \alpha_2), \\
    \alpha_0 - \alpha_3 &\rightarrow \beta(\alpha_0 - \alpha_3),
    \label{eq:1q_typeI_spoof_transform}
\end{split}
\end{equation}
which indicates that \textit{a change occurs to all four coefficients} in such a way that it is not detectable by a set of compatible measurements -- in this case the computational basis. Thus, the crucial question is whether the channel $\mathcal{E}'$ preserves the diagonal of density matrices, as these are the elements where any changes would be detected by computational basis measurements. We can show that the transformation~(\ref{eq:1q_typeI_spoof_transform}) causes no such change via the natural representation, where the action on the vectorised density matrix Eq.~(\ref{eq:naturalchannelrep}) reads
\begin{equation}
\begin{split}
    &\begin{pmatrix}
        \alpha_0 + \alpha_3 & 0 & 0 & \beta(\alpha_0 - \alpha_3) \\
        0 & \alpha_1 + \alpha_1 & \beta(\alpha_1 - \alpha_2) & 0 \\
        0 & \beta(\alpha_1 - \alpha_2) & \alpha_1 + \alpha_2 & 0 \\
        \beta(\alpha_0 - \alpha_3) & 0 & 0 & \alpha_0 + \alpha_3
    \end{pmatrix}
    \begin{pmatrix}
        \rho_{00} \\
        \rho_{01} \\
        \rho_{10} \\
        \rho_{11} \\
    \end{pmatrix} 
    \\
     & =\begin{pmatrix}
        (\alpha_0 + \alpha_3)\rho_{00} + (\alpha_1 + \alpha_2)\rho_{11}\\
        \beta(\alpha_0 - \alpha_3)\rho_{01} +  \beta(\alpha_1 - \alpha_2)\rho_{10}\\
        \beta(\alpha_0 - \alpha_3)\rho_{10} +  \beta(\alpha_1 - \alpha_2)\rho_{01} \\
        (\alpha_0 + \alpha_3)\rho_{11} + (\alpha_1 + \alpha_2)\rho_{00} \\
    \end{pmatrix}. 
\end{split}
\end{equation}

This shows explicitly how the transformations in Eq.~(\ref{eq:1q_typeI_spoof_transform}) cannot change the diagonal elements (first and last) of the vectorised density matrix, despite the fact that all the coefficients of the Pauli channel do change. Thus these transformations form a gauge freedom with respect to the set of compatible measurements . Geometrically, this gauge degree of freedom corresponds to a line contained in the tetrahedron representing all valid 1-qubit Pauli channels. See Sec.~\ref{sec:one_qubit_geom} and Fig.~\ref{fig:1Q_pauli_vis} for further discussion.

\subsubsection*{Type-II Spoofing}
Next, we consider Type-II spoofing detailed in Sec.~\ref{sec:TypeIISpoofing}, where we find a spoofing channel which is not necessarily divisible per Fig.~\ref{fig:spoofers}. From Theorem \ref{thm:type_II_spoofing}, we see that the Choi matrix is of the form
\begin{equation}
    J_{\mathcal{E}'} = 
    \begin{pmatrix}
        \alpha_0 + \alpha_3 & 0 & 0 & \gamma \\
        0 & \alpha_1 + \alpha_2 & \beta & 0 \\
        0 & \beta & \alpha_1 + \alpha_2 & 0 \\
        \gamma & 0 & 0 & \alpha_0 + \alpha_3
    \end{pmatrix},
\end{equation}
where $\beta, \gamma \in \mathbb{R}^+$, so that the spoofing channel is also a Pauli channel. This induces the following change in the coefficients $\{\alpha_j\}$ of the channel,
\begin{equation}
    \begin{split}
    \alpha_1 - \alpha_2 &\rightarrow \beta, \\
    \alpha_0 - \alpha_3 &\rightarrow \gamma.
    \label{eq:1q_typeII_spoof_transform}
\end{split}
\end{equation}
Notice that this transformation leaves the sum $\alpha_0 + \alpha_3$ and $\alpha_1 + \alpha_2$ unchanged, and is equivalent to,
\begin{equation}
\label{eq:better_way_alpha_changes}
    \begin{split}
        \alpha_0' &= \alpha_0 + \gamma/2,\\
        \alpha_1' &= \alpha_1 + \beta/2, \\
        \alpha_2' &= \alpha_2 - \beta/2, \\
        \alpha_3' &= \alpha_3 - \gamma/2.
    \end{split}
\end{equation}
Again, counter to the naive intuition, this spoofing channel changes \textit{all four} of the coefficients $\{\alpha_j\}$ despite not being detectable in the computational basis.

\subsection{Geometric Comparison of One Qubit Channel Spoofing}
\label{sec:one_qubit_geom}
\begin{figure}[H]
  \centering
  \input{figures/labels_spoofer_geometry}
  \vspace{-5em}
  \caption{Comparison between Type-I and Type-II spoofers on one-qubit Pauli channels in terms of the degrees of freedom. The vertices of the tetrahedron correspond to unitary maps involving only one of the four available Pauli operators. The red (solid) line represents the single gauge degree of freedom in Type-I Spoofing, whilst the clipped blue plane represents the two degrees of freedom in Type-II Spoofing. In this form, we observe that every Type-II Spoofer is a Type-I Spoofer and not vice versa since the red dashed line is embedded in the plane. In this figure, the original Pauli channel has coefficients $\{\alpha_0,\alpha_1,\alpha_2,\alpha_2\} = \{0.1,0.1,0.1,0.7\}$.}
  \label{fig:1Q_pauli_vis}
\end{figure}
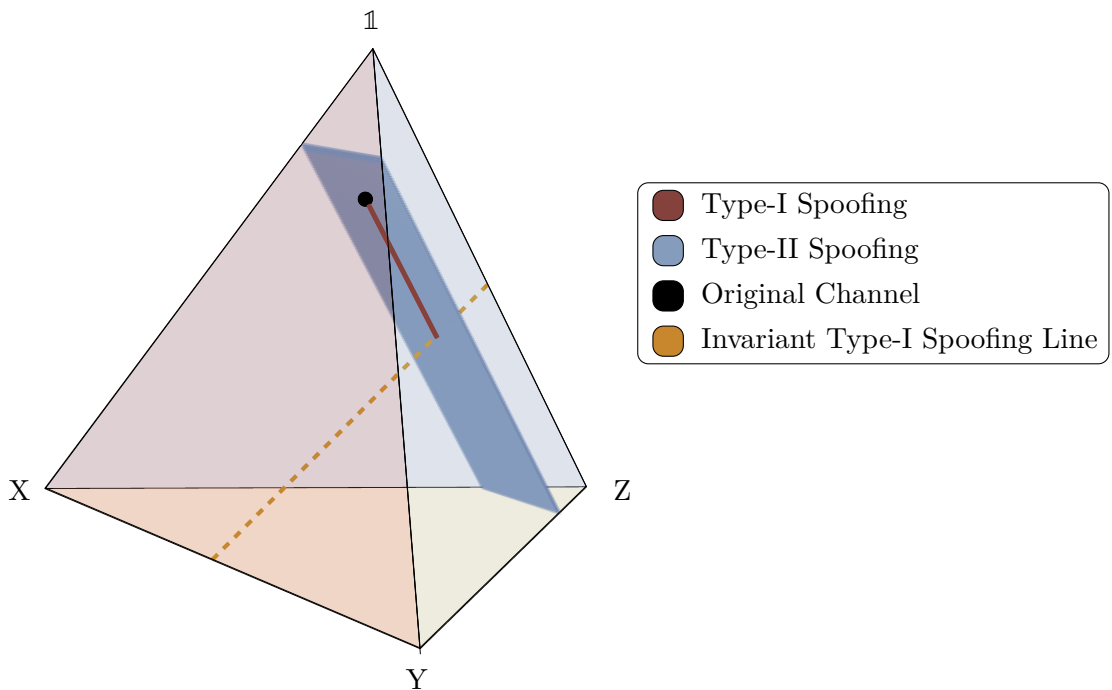

\noindent To compare these two types of spoofing on one-qubit Pauli channels, observe that Type-I spoofing has a single gauge freedom, $\{\beta\}$, whilst Type-II spoofing has two, $\{\beta, \gamma\}$. The fact that there are two gauge degrees of freedom in Type-II spoofing can be visualised in the space of Pauli maps for 1 qubit, i.e., a tetrahedron, $\mathcal{T}$, as shown in Fig.~\ref{fig:1Q_pauli_vis}. 

\onecolumngrid

\twocolumngrid

The two degrees of freedom means the family of Type-II spoofing channels spans a clipped-plane contained within the tetrahedron, 
$\text{span}(\beta, \gamma) \cong \mathbb{R}^2\cap \mathcal{T}$. On the other hand, Type-I spoofing channels form a line linking the original channel to the line joining two channels, $(\mathbb{I}\rho\mathbb{I} + Z \rho Z )/2$ and $(X \rho X + Y \rho Y)/2$. 
We can see what is so special about this line -- every channel on the line is invariant under the Type-I spoofer, i.e., this family of channels does not change by varying $\beta$.
Of note here is the fact that the Type-I line is contained within the clipped plane, meaning Type-II spoofing forms a superset of the family of spoofing channels of Type-I. Operationally, \textit{every Type-I spoofer is also a Type-II, but not every Type-II spoofer is a Type-I spoofer}. This leads us to the following corollary:

\begin{corollary}
    Type-II spoofers form a superset of Type-I spoofers, however Type-I spoofers can be applied universally with no knowledge of the original channel.
\end{corollary}

\noindent The reason Type-I spoofing requires no prior knowledge is because it concatenates with the original channel $\mathcal{E}$, and its structure has no dependence on the original coefficients of the channel. We only required that the concatenation acts as the identity channel along detectable components in the natural representation. This is in contrast with Type-II spoofing, where the new Choi matrix of the spoofing channel $J_{\mathcal{E}'}$ explicitly contains diagonal blocks whose elements take the same value as the original channel. Thus Type-II spoofing requires prior knowledge.

\section{Discussion}
\label{sec:discussion}
In principle, quantum mechanics allows for the simultaneous measurement in any commuting basis.
However, in practise, making measurements different from the computational basis is done by applying
a pre-measurement local transformation \cite{huang2020predicting, garcia2021learning}.  Provided a spoofer has knowledge of the chosen measurement basis ahead of compiling, this means spoofing any projective measurement basis is possible. This is the case in cloud-based quantum hardware, where providers know both the channel and the measurement basis before executing the channel.

Consequently, an important strategy for device certification therefore arises in this setting. Namely, one can only certify that a channel belongs to a 
certain spoofing equivalence class (and not certify a unique channel) due to the fact that spoofers would go undetected with projective measurements alone. 
Moreover, within a given equivalence class, the Kraus rank can change, which affects the simulation hardness. This in turn impacts the quantum advantage that can be certified on cloud-based quantum hardware. 
One way we could perform device certification in lieu of spoofing is therefore to find the \textit{minimal} Kraus rank of a spoofing class of a given channel, see Sec.~\ref{sec:sinkhorn}. 
Notice that the minimal Kraus rank channel is itself \textit{not unique}. This is because the number of linearly independent degrees of freedom was $\mathcal{O}(d^4)$, but constraining the rank of spoofing channel to be of minimal rank, $d$, introduces a further $d^2 - d \leq d^4$ constraints for the spoofing degrees of freedom. In fact, the only time the minimal-rank spoofing channel is unique is when
\begin{align}
    (d^4 - d^3 - d^2 + d) - (d^2 - d) &= 0 \\
    d^2(d -2)(d+1) &=0
\end{align}
so uniqueness of the minimal rank channel is only true when $d = 2$.

As briefly mentioned in the introduction, one might argue that techniques like classical shadows~\cite{Huang_2020} (introducing classical randomness into measurements) mean 
that we can perform informationally complete measurements.
In this sense, an inherent amount of classical randomness could protect against spoofing. This is possible for an
adversary when we consider that adding classical randomness is within the bounds of mixed unitary channels in the first place. However, even with classical randomness, we would only be able to detect a spoofer when given access to exponential resources. This is because we need to estimate the full quantum channel to detect a spoofer, requiring a large dataset which is exponential in the number of bodies; in fact, distinguishing
quantum channels is a \texttt{QIP}-hard problem \cite{rosgen2009computational}.  

For a small system, our spoofer would therefore easily
be detected. But with the promise of quantum devices to scale to many hundreds of qubits, acquiring sufficient
amounts of data to fully characterise a quantum channel is prohibitively expensive.
This is what we may exploit in order to design spoofing strategies.  A spoofer can go undetected when the dataset used to characterise a channel is sub-exponential in the number of shots. 
However, the spoofer will only go undetected with some success probability, as the randomness of the 
shadow-based measurements means that certain changes could be identified. 
For example, a user who wishes to certify a channel on $N$-qubit system with $M$ shots will detect the spoofer with probability $p \sim \mathcal{O}(M/4^N \epsilon^2)$, for a precision $\epsilon$, as highlighted in Sec.~\ref{sec:PauliCaseStudy}.

One might argue that a valid defense strategy against this scenario is to attempt to characterise a quantum channel's Choi matrix to an accuracy of $\epsilon$. Whilst this requires exponential resources in general, much attention has been given to trying to lower this complexity overhead. For example, recent works such as \cite{torlai2023quantum} have used a tensor network representation of the Choi matrix estimator $\hat{J_{\mathcal{E}}}$, meaning that their search space forms a polynomially-scaling sub-manifold of the space of channels. However, tensor networks suffer from the so-called entanglement area law \cite{calabrese2004entanglement},
meaning they cannot be used to certify channels with higher entropy than permitted by this approximation. From an adversarial perspective, this leaves a spoofer with the option to vary elements of a Choi matrix that are not captured by this representation. In other words, by changing the channel such that its reduction under these changes to a tensor network representation is undetectable, an adversary can still cause a user to falsely certify a quantum channel's Choi matrix.
In particular, we have numerically verified that, for all 1-qubit Pauli channels, one can find an entanglement-breaking~\cite{Horodecki_2003} channel in its spoofing class. 
This means that, in the scenario of sending an entangled particle to a receiver, it is necessary to use exponential number of shots in the randomized setting to prevent a manipulation by a malicious adversary.

\subsection{Spoofers with Minimal Kraus Rank}
\label{sec:min_kraus_rank_importance}

In the above, we alluded to the complications that arise when attempting to certify a quantum channel in cloud-based settings due to the fact that spoofing equivalence classes contain channels with differing Kraus rank. In this subsection, we highlight the importance of being able to determine the \textit{minimum} Kraus rank of a given equivalence class for quantum advantage, quantum simulation, and variational quantum channels.

Consider the scenario where a user wishes to certify the quantum advantage of a device being managed by an untrusted adversary, for example in cloud-based quantum hardware. In order to certify that their device provides quantum advantage, they may choose to estimate the simulation hardness of the device's quantum channel representation, using a measure such as the Kraus rank (i.e., the number of non-zero eigenvalues of the channel's Choi matrix). In the case where a user has access only to projective measurements in a specified basis, the adversary providing the channel is free to vary the other parameters of the Choi matrix that define its equivalence class. Therefore a user must know the minimal Kraus rank of their quantum process, or risk being mislead about its resource requirements. This is because the Kraus rank of the members of the spoofer's equivalence class can range from $d$ to $d^2$ for a $d$-dimensional quantum channel, 
whilst the Choi rank of a channel is indicative of the resources necessary to simulate it \cite{watrous2018theory}.

Since the adversarial hardware provider will know both the channel \textit{and} measurements before compiling, it is possible for the adversary to simulate a user's quantum process in such a way that the actual channel being run on the device has a lower Kraus rank that the one a user intended. Furthermore, since these changes leave the outcome distributions of the user's measurements unchanged, any changes made by the adversarial provider are undetectable by the user. It is therefore of critical importance for users to be able to find the channel realizing minimum Kraus rank inside a given equivalence class in order to defend against this adversarial strategy.

Conversely, an adversary that wishes to spoof the original channel needs to know with what channel they can do it most efficiently.
Both of these applications, albeit from opposite directions, underline the importance of the subsequent study. 

\section{Finding Spoofers with Minimal Kraus Rank}
\label{sec:low_rank_sims}
\noindent Having found the structure of the spoofing families in Section \ref{sec:parametrisation_of_spoofing}, we are in a position to solve the practical problem of minimising the Kraus rank of a given channel $\mathcal{E}$ (for the discussion of its importance, see Sec.~\ref{sec:discussion}). 
Specifically, we seek a channel $\mathcal{E}'$ belonging to the Type-II spoofing family of the original channel $\mathcal{E} \sim_{\scalebox{0.5}{$\mathcal{M}$}} \mathcal{E}'$, such that the Kraus rank of $\mathcal{E}'$ is the smallest of its equivalence class defined by Eq.~(\ref{eq:equivalence_relation}).

We can numerically tackle this problem via a Sinkhorn-like algorithm \cite{sinkhorn1967concerning}. Our algorithm relies on singular value decomposition, and can reduce the Kraus rank from a generic $d^2$ to $d$, as shown in Sec~\ref{sec:sinkhorn}. We also provide an analytic algorithm to minimize the Kraus rank of a given channel to its minimum bound for the special case of Pauli channels in Sec~\ref{sec:analytic_pauli_lowering}.

For a generic channel, we note that no further lowering is possible, as the rank of the Choi matrix cannot be smaller than the rank of its sub-matrices. 
Therefore, it suffices that at least one of $d \!\times\! d$ diagonal blocks of the Choi matrix of the channel is of full rank to restrict the lowest possible Kraus rank of any channel in its spoofing equivalence class to $d$. 
Given that both our algorithms achieve this value, we conclude that they fully solve the rank problem for spoofing of channels. A \texttt{Mathematica} implementation of the Sinkhorn-like algorithm can be found at GitHub~\cite{github}.
\subsection{Sinkhorn-like Algorithm for an Arbitrary Channel}
\label{sec:sinkhorn}
This numerical algorithm uses an analogous reasoning to the Sinkhorn algorithm \cite{sinkhorn1967concerning}, which was first use to generate bistochastic matrices. We start with a $d$-dimensional quantum channel, $\mathcal{E}$, generically of Kraus rank $d^2$. 
Our aim it find a channel in the same spoofing equivalence class which achieves the minimal Kraus rank $d$. 
We start out by finding the closest matrix via a low-rank approximation~\cite{Schmidt_1907,Eckart_1936}. 
However, in general this closest matrix will not be a valid Choi matrix.
Therefore, to properly lower the rank we need to take into account two factors: (i) the usual conditions for the Choi matrix to be a CPTP map, and (ii) we must ensure that diagonal blocks of its Choi matrix are equal to the ones of the original channel, per Theorem.~\ref{thm:type_II_spoofing}.

To simultaneously satisfy these two factors, we may alternate between them, as shown in Algorithm~\ref{alg:sinkhorn}. In the algorithm, by $\{\lambda_i\}_{i=1}^{d^2}$ we refer to the eigenvalues of the current Choi matrix in decreasing order, while $[J_{\mathcal{E}'}]_{ij}$ denotes the $\{i,j\}$-th block of dimension $d\!\times\! d$ of the matrix $J_{\mathcal{E}'}$.

\begin{algorithm}[H]
	\caption{Sinkhorn-like lowering the Kraus rank of a channel}\label{alg:sinkhorn}
	\begin{algorithmic}[1]
            \State \textbf{Input:} Choi matrix, $J_{\mathcal{E}}$, of the original channel $\mathcal{E}$ of rank up to $d^2$.
            \State $J_{{\mathcal{E}'}} \leftarrow J_{{\mathcal{E}}}$
		\While {$\lambda_{d+1} > \epsilon$}
                
			\State $J_{\mathcal{E}'} = U D U^\dagger$
                \Comment{ Find the low-rank approximation to Choi matrix $J_{\mathcal{E}'}$ via singular value decomposition}                
                \State $J_{\mathcal{E}'}\leftarrow U_d D_d U^\dagger_d$ \Comment{ Truncate the matrices by discarding all but $d$ largest singular values}
                \State Set the diagonal blocks of $J_{\mathcal{E}'}$ to be equal to the original $J_{\mathcal{E}}$
                \For{$i \in \{1,...,d\}$}
                    \State $[J_{\mathcal{E}'}]_i\leftarrow [J_\mathcal{E}]_i$
                \EndFor
                \For{$i, j \in \{1,...,d\}, \, i \neq j$}
                    \State $[J_{\mathcal{E}'}]_{ij}\leftarrow [J_{\mathcal{E}'}]_{ij} - \mathbb{I}\, \text{Tr}([J_{\mathcal{E}'}]_{ij})/d $
                    \Comment{Set the off-diagonal blocks to be traceless}
                \EndFor
                \State $J_{\mathcal{E}'} \leftarrow \frac{1}{2}(J_{\mathcal{E}'} + J^\dagger_{\mathcal{E}'} ) $
                \Comment{ Make the Choi matrix Hermitian for numeical stability}
            \EndWhile
            \State \textbf{Output:} Choi matrix $\mathcal{J}_{\mathcal{E}'}$ of rank $d$ in the same equivalence class as $\mathcal{E}$ defined by Eq.~(\ref{eq:equivalence_relation}).
	\end{algorithmic} 
\end{algorithm}

The algorithm produces the spoofing channel, $\mathcal{E}'$, with minimal Kraus rank $d$, in its Type-II spoofing equivalence class, in which the original channel, $\mathcal{E}$, is also a member. That is, by using $\mathcal{E}'$, we reduce the Kraus rank from $d^2$ (in the generic case) to $d$ whilst maintaining the exact outcome distributions for a projective measurement, per Eq.~(\ref{eq:equivalence_relation}). 
All of the initial channels, $\mathcal{E}$, that we have tested through Algorithm~\ref{alg:sinkhorn} have produced a channel of the minimal rank in their equivalence class, suggesting that the algorithm may work generally for any input channel.
Importantly, the convergence rate of the eigenvalues is exponential, as shown in Fig.~\ref{fig:sinkhorn} for an exemplary channel of dimension $20$ and $50$.
We observed the complexity growth that scales with the dimension of the system as $d^4$, reflecting the growing number of parameters that describe a channel.

\onecolumngrid
\begin{center}
\begin{figure}[H]
    \centering
    \input{TikZ/eigenvalues_convergence}
    \caption{The convergence of eigenvalues of the Choi matrix of a generic (a) $d = 20$ and (b) $d = 50$ dimensional quantum channel under Sinkhorn-like algorithm \cite{sinkhorn1967concerning}. In the upper panel, out of $20^2 = 400$ eigenvalues, $20$ are positive, denoted by $\{1,...,20\}$ in the figure, while the rest exponentially converge to zero, similarly in the lower panel. By setting a tolerance value, we can therefore set these exponentially vanishing eigenvalues to zero, finding the Type-II spoofing matrix for the projective measurement set defined in Sec.~\ref{sec:statement_of_problem}.}
    \label{fig:sinkhorn}
\end{figure}
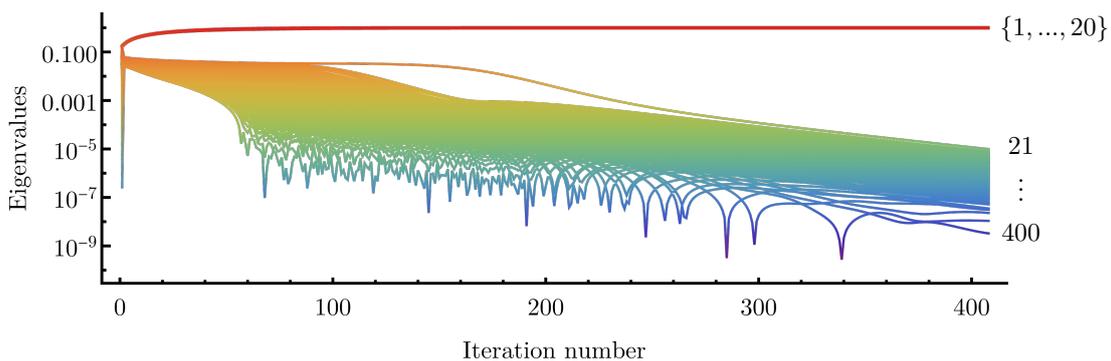
\end{center}
\twocolumngrid

\subsection{Analytical Spoofing Algorithm for Pauli Channels}
\label{sec:analytic_pauli_lowering}
The algorithm detailed above may be universal in the sense that it works for any arbitrary channel in $d$-dimensions. However, it runs numerically, which can be cumbersome as the system size grows.
Here, we propose an algorithm for lowering the rank that is specifically tailored to Pauli channels, with the added advantage of being analytical. It shows that any $N$-qubit Pauli channel's projective outcome distributions can be simulated by a quantum channel with Kraus rank $d$.

Let us start with a simple observation concerning Choi matrix of a Pauli channel on $N$ qubits, defined by its Kraus operators $\{\sqrt{\alpha_i} P_i\}_i$. The location of non-zero elements for each of the $4^N$ rows in the Choi matrix forms a pattern which is repeated $2^N$ times. To see how this works, we consider the example shown Fig.~\ref{fig:non-zero_pattern} for 2- and 3-qubit channels, which shows that the same pattern of non-zero elements repeats itself four and eight times respectively for 2- and 3-qubit channels.

A given Kraus coefficient of the Pauli channel, $\sqrt{\alpha_i}$, corresponds to a unique non-zero pattern which is repeating.
For each of the sets, we may take the full set of the corresponding coefficients $\{\alpha_i\}_{i \in S}$ and sum them, obtaining $\beta_S = \sum_{i \in S} \alpha_i$.
Then, by setting one of these coefficients to be equal to $\beta_S$ and the rest to 0, we do not leave the Type-II spoofing equivalence class of the original channel.
Consequently, the Kraus rank of the channel lowers quadratically, from generic $4^N$ to $2^N$. This result is analogous to Algorithm~\ref{alg:sinkhorn} from the previous subsection, but works analytically.
An exemplary usage of this algorithm for the case of 1-qubit channels is provided in App.~\ref{app:1qubit}.
The advantage of this algorithm comes from engineering linear dependence via an analytical change of the free parameters.  
For further explanation of this algorithm we refer the reader to our Github code~\cite{github}.

\onecolumngrid
\begin{center}
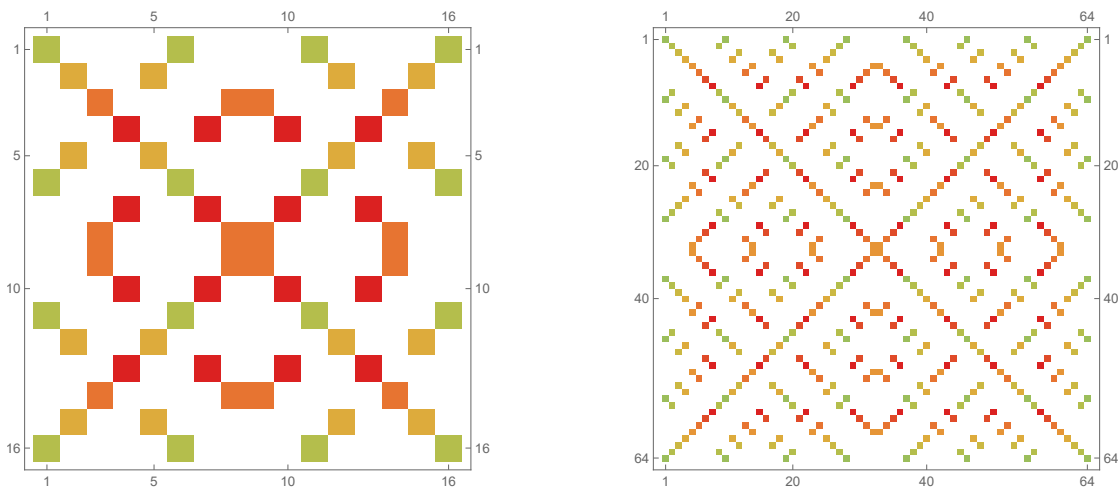
\begin{figure}[H]
    \centering
        \subfloat{\input{TikZ/non-zero_pattern_2q}}
        \hspace*{\fill}
        \subfloat{\input{TikZ/non-zero_pattern_3q}} 
    \caption{The non-zero patterns for 2-qubit (left panel) and 3-qubit (right panel) Choi matrices for Pauli channels, where the colors distinguish between different patterns. Observe that, for example, in the left panel rows: 1, 6, 11, and 16 share the same pattern of non-zero elements. In total, there are $2^N = 4$ and $8$ different patterns respectively for $2$-qubit and $3$-qubit channels.}
    \label{fig:non-zero_pattern}
\end{figure}
\end{center}
\twocolumngrid

\section{Concluding Remarks}
\label{sec:conclusion}
\noindent One of the most important problems in characterising quantum devices is determining the quantum channel representation of their evolution. 
In this contribution, we have first illustrated a key consequence of the \texttt{QIP}-hardness of quantum channel discrimination \cite{rosgen2009computational}. Namely, one cannot establish a direct link (i.e., with full certainty) between outcome statistics and
Choi matrices for quantum channels with sub-exponential resources. We have laid grounds, based on this complexity, for spoofing an arbitrary quantum channel in $d$-dimensional Hilbert space in such a way 
that its Kraus rank can be lowered from generic $d^2$ to $d$. This can be done in an undetectable way for fixed measurement bases,
while the success probability of detection under shadow-based measurements scales as $\mathcal{O}(M/d^2)$.
Given the Choi matrix, $J_{\mathcal{E}}$, of any quantum channel, $\mathcal{E}$, we showed how to find this minimum via a Sinkhorn-like algorithm, whose efficacy was numerically demonstrated for quantum channels up to dimension $50$.
For the case of Pauli channels, the same improvement for the minimal Kraus rank was found analytically in a single step by engineering linear dependence of rows in the Choi representation.
Our findings open new avenues for efficient quantum channel simulation and adversarial spoofing strategies.

Given our findings, one might wonder how the number of gauge degrees of freedom changes when we allow for more than one projective measurement. Specifically, we can refine the equivalence classes by appending more bases to the permitted measurement set $\mathcal{M}$
from Eq.~(\ref{eq:sameProbDiffChannel}). For example, if we allow
\begin{equation}
    \mathcal{M} =
\{\ket{q}\!\!\bra{q},
U_1\ket{q}\!\!\bra{q}U_1^{\dagger},\ldots, U_{k-1}\ket{q}\!\!\bra{q}U_{k-1}^{\dagger}\},
\end{equation}
we could establish a minimal Kraus rank which goes from $d^2 \rightarrow f(k) d$ under this larger set of projective measurements, for some $f(k)$ to be determined. Another future direction is to determine how an equivalence class may be structured when we instead require that the spoofing channel has the same outcome statistics up to a given polynomial order in the moments of outcome distributions. This would re-frame the spoofing families' equivalence classes to be effective for any projective measurement for finite statistics. We leave these suggestions as open problems for the quantum information community.

\subsection*{Acknowledgements}

We would like to thank Jakub Czartowski, Jaros{\l}aw Korbicz, Marcin P{\l}odzie\'n, Jos\'e Ram\'on Mart{\'i}nez, Antonio Ac{\'i}n,  and Karol \.Zyczkowski for fruitful discussions and comments.

TH was supported by the Government of Spain (Severo Ochoa CEX2019-000910-S, Quantum in Spain, FUNQIP and European Union NextGenerationEU PRTR-C17.I1), the European Union (PASQuanS2.1, 101113690 and Quantera Veriqtas), Fundació Cellex, Fundació Mir-Puig, Generalitat de Catalunya (CERCA program), the ERC AdG CERQUTE and the AXA Chair in Quantum Information Science.
GRM acknowledges funding from the European Innovation Council accelerator grant COMFTQUA, no. 190183782.

\bibliography{references}

\appendix

\section{Matrix Reshuffling}\label{app:reshuffling}
We can switch to the Choi representation from the natural representation via a reshuffling operation, $S_{\mathcal{E}} \overset{R}{\mapsto} J_{\mathcal{E}}$. 
The simplest way to understand reshuffling is through double indices 
\begin{equation}
    M_{ij,kl}^R = M_{ik,jl}.
\end{equation}
In terms of the matrix-form of Eq.~(\ref{eq:naturalchannelrep}), this reshuffling is enacted by  
\begin{equation}
\begin{split}
    &S_{ijkl} \mapsto J_{\mu \nu} \;\; \text{s.t.} \\
    &(i,j) \mapsto \mu  = d(i-1) + k, \\ 
    &(k,l) \mapsto \nu = d(j-1) + l.
\end{split}
\label{eq:nat_to_choi}
\end{equation}

\section{Conditions to Simulate a Channel in Kraus Representation}
\label{app:conditions_in_kraus}
To simultaneously satisfy Eq.~(\ref{eq:sameProbDiffChannel}) and Eq.~(\ref{eq:diffChoi}), we require two types of conditions for the Kraus operators~\footnote{For simplicity, we assume that the number of Kraus operators is the same. The rationale in the other case is basically the same. For the latter part of this appendix, we remove this assumption and work in the more general scenario.}:
\begin{itemize}
    \item \textbf{Condition 1} -- for all input states and measurement outcomes,
    \begin{equation}
    \sum_{r = 0}^R \sum_{m,n = 0}^{d - 1}K^{(r)}_{qn}\rho_{nm}K^{(r)\dagger}_{mq} = \sum_{r = 0}^R \sum_{m,n = 0}^{d - 1}K'^{(r)}_{qn}\rho_{nm}K'^{(r)\dagger}_{mq},
    \label{eq:condition1}
    \end{equation}
    where $K_{qn}^{(r)}$ (resp. $K_{qn}'^{(r)}$)  form a Kraus representation of the channel $\mathcal{E}$ (resp. $\mathcal{E}'$).
    \item \textbf{Condition 2} -- there exist values for the indices $\{i,j,k,l\} \in \{0,\ldots,d-1\}$ such that 
    \begin{equation}
    \sum_{r = 0}^{R} \sum_{l,n = 0}^{d-1}\left(K^{(r)}_{ki}K^{(r)\dagger}_{nj} - K'^{(r)}_{ki}K'^{(r)\dagger}_{nj}\right) \neq 0.
    \label{eq:condition2}
\end{equation}
\end{itemize}

\begin{proof}
Eq.~(\ref{eq:condition1}) is simply a restatement of Eq.~(\ref{eq:sameProbDiffChannel}) at the level of Kraus operators. To see why this is the case, consider calculating the outcome probability 
\begin{equation}
    p_{\mathcal{E} }(q|\rho) = \text{Tr}(\mathcal{E} (\rho) \ket{q}\!\!\bra{q}).
\end{equation}
As introduced above, let the POVM set be computational basis measurements.
In this basis, the Kraus operators of $\mathcal{E}$ are,
\begin{equation}
    K^{(r)} = \sum_{i,j = 0}^{d - 1} K^{(r)}_{i,j} \ket{i}\!\!\bra{j}, 
\end{equation}
whilst the state, $\rho  = \sum_{n,m} \rho_{nm}\ket{n}\!\!\bra{m}$. 
Using this basis, we can express the action of $\mathcal{E}$ as,
\begin{equation}
\begin{split}
    \mathcal{E} (\rho) &= \sum_{r = 0}^R \sum_{i,j,k,l,m,n = 0}^{d - 1} K^{(r)}_{ij}\ket{i}\!\!\bra{j} \rho_{nm} \ket{n}\!\!\bra{m} K^{(r)\dagger}_{kl} \ket{l}\!\!\bra{k}. \\
    &= \sum_{r = 0}^R \sum_{i,l,m,n = 0}^{d - 1}K^{(r)}_{in}\rho_{nm}K^{(r)\dagger}_{ml}\ket{i}\!\!\bra{l}.
\end{split}
\end{equation}
This allows us to calculate the outcome probabilities
\begin{widetext}
\begin{equation}
    \begin{split}
        \text{Tr}(\mathcal{E} (\rho) \ket{q}\!\!\bra{q}) &= \text{Tr}\left( 
        \sum_{r = 0}^R \sum_{i,l,m,n = 0}^{d - 1}K^{(r)}_{in}\rho_{nm}K^{(r)\dagger}_{ml}\ket{i}\!\braket{l|q}\!\bra{q}
        \right) \\
        &= \sum_{r = 0}^R \sum_{i,l,m,n = 0}^{d - 1}K^{(r)}_{in}\rho_{nm}K^{(r)\dagger}_{ml}\text{Tr}\big(\ket{i}\!\braket{l|q}\!\bra{q}\big) \\
        &= \sum_{r = 0}^R \sum_{m,n = 0}^{d - 1}K^{(r)}_{qn}\rho_{nm}K^{(r)\dagger}_{mq}.
    \end{split}
\end{equation}
\end{widetext}

Let us understand this form a little better. First, this condition is over the \textit{sum} of Kraus operators. Thus, each Kraus matrix (or any subset thereof) can be different provided the \textit{collective} action on the state $\rho$ is the same. Second, both the left-hand side and the right-hand side have $R + 1$ Kraus operators, but in general this is not necessary since they are both summed over. This shows why a spoofing channel can in principle have a lower rank than the original channel, provided each term in the sum contributes on average a larger amount to this probability. 


We can summarise this by requiring
\begin{widetext}
    \begin{equation}
    \sum_{r = 0}^R \sum_{m,n = 0}^{d - 1}K^{(r)}_{qn}\rho_{nm}K^{(r)\dagger}_{mq} = \sum_{r = 0}^{R'} \sum_{m,n = 0}^{d - 1}K'^{(r)}_{qn}\rho_{nm}K'^{(r)\dagger}_{mq}, \; \; R' \leq R, \; \; \forall \; q \in \{0, d-1\},
    \label{eq:sameProbAllQ}
\end{equation}
\end{widetext}
which takes into account the above points. Hence, we see that demanding the outcome probabilities be the same irrespective of the input state is simply requiring Eq.~(\ref{eq:condition1}) at the level of Kraus operators. 

The second condition we require is that the Choi matrices be different. Eq.~(\ref{eq:condition2}) is simply a statement of this at the level of Kraus operators. Recall that the Choi matrix of a channel $\mathcal{E} $ 
can be expressed in the computational basis via Kraus operators,
\begin{equation}
    \begin{split}
        J_{\mathcal{E} } &= (\mathbb{I} \otimes \mathcal{E} ) \sum_{ij = 0}^{d- 1} \ket{i}\!\!\bra{j} \otimes \ket{i}\!\! \bra{j}\\
        &= \sum_{kij = 0}^{d-1}\ket{k} \braket{k}{i} \bra{j} \otimes \mathcal{E} (\ket{i}\!\!\bra{j}) \\
        &= \sum_{ij = 0}^{d-1} \left(\ket{i}\!\!\bra{j}  \otimes \sum_{r = 0}^{R}\sum_{klmn = 0}^{d-1}K^{(r)}_{kl}\ket{k}\braket{l}{i}\braket{j}{m}\bra{n}K^{(r)\dagger}_{nm}\right) \\
        &= \sum_{ij = 0}^{d-1} \left(\ket{i}\!\!\bra{j}  \otimes \sum_{r = 0}^{R} \sum_{ln = 0}^{d-1}K^{(r)}_{ki}K^{(r)\dagger}_{nj} \ket{k}\!\!\bra{n} \right) \\
        &= \sum_{r = 0}^{R}\sum_{ijln = 0}^{d-1} \ket{i}\!\!\bra{j} \otimes K^{(r)}_{ki}K^{(r)\dagger}_{nj} \ket{k}\!\!\bra{n}.
    \end{split}
\end{equation}
Hence, the condition that the two Choi matrices, $J_{\mathcal{E} }$ and $J_{\mathcal{E}' }$, be different can be expressed as requiring the \textit{difference} between the two to be non-zero. 
Assuming the same number of Kraus matrices for both channels, we require
\begin{widetext}
\begin{equation}
    \sum_{r = 0}^{R}\sum_{ijln = 0}^{d-1} \ket{i}\!\!\bra{j} \otimes K^{(r)}_{ki}K^{(r)\dagger}_{nj} \ket{k}\!\!\bra{n} - \sum_{r = 0}^{R}\sum_{ijln = 0}^{d-1} \ket{i}\!\!\bra{j} \otimes K'^{(r)}_{ki}K'^{(r)\dagger}_{nj} \ket{k}\!\!\bra{n} \neq \mathbb{0},
\end{equation}
\end{widetext}
where $\mathbb{0}$ is the matrix of 0s in dimension $d^2$. By factorising the left-hand side of the tensor product, this expression can be simplified to
\begin{widetext}
    \begin{equation}
    \sum_{ij = 0}^{d-1} \ket{i}\!\!\bra{j}  \otimes \sum_{r = 0}^{R} \sum_{ln = 0}^{d-1}\left(K^{(r)}_{ki}K^{(r)\dagger}_{nj} - K'^{(r)}_{ki}K'^{(r)\dagger}_{nj}\right)\ket{k}\!\!\bra{n} \neq \mathbb{0}.
\end{equation}
\end{widetext}
Since the LHS of the tensor product has coefficients $1$ for all $i,j$, we see that the non-zero requirement is trivially satisfied in this space. Hence, focusing our attention on the RHS of the tensor product, we require for \textit{at least} a single pair of indices $i,j$ that
\begin{equation}
    \sum_{r = 0}^{R} \sum_{ln = 0}^{d-1}\left(K^{(r)}_{ki}K^{(r)\dagger}_{nj} - K'^{(r)}_{ki}K'^{(r)\dagger}_{nj}\right) \neq 0.
\end{equation}
This concludes the proof, as this is the result we stated in Eq.~(\ref{eq:condition2}), when the number of Kraus matrices is the same.
\end{proof}

On the other hand, if the two channels have different number of Kraus matrices, $R+1$, $R'+1$, 
then we require
\begin{widetext}
    \begin{equation}
    \sum_{r = 0}^{R}\sum_{ijln = 0}^{d-1} \ket{i}\bra{j} \otimes K^{(r)}_{ki}K^{(r)\dagger}_{nj} \ket{k}\bra{n} - \sum_{r = 0}^{R'}\sum_{ijln = 0}^{d-1} \ket{i}\bra{j} \otimes K'^{(r)}_{ki}K'^{(r)\dagger}_{nj} \ket{k}\bra{n} \neq \mathbb{0},
\end{equation}
\begin{equation}
    \implies \sum_{ij = 0}^{d-1} \ket{i}\!\!\bra{j} \otimes 
    \left( 
    \sum_{r = 0}^{R}\sum_{kn = 0}^{d-1}K^{(r)}_{ki} K^{(r)\dagger}_{jn} \ket{k}\!\!\bra{n} - \sum_{r = 0}^{R'}\sum_{kn = 0}^{d-1}K^{(r)}_{ki} K^{(r)\dagger}_{jn} \ket{k}\!\!\bra{n}
    \right) \neq \mathbb{0},
\end{equation}
\begin{equation}
    \implies \sum_{ij = 0}^{d-1} \ket{i}\!\!\bra{j} \otimes 
    \sum_{kn = 0}^{d-1} \ket{k}\!\!\bra{n}
    \left(
    \sum_{r = 0}^{R} K^{(r)}_{ki} K^{(r)\dagger}_{jn} - \sum_{r = 0}^{R'}K'^{(r)}_{ki} K'^{(r)\dagger}_{jn}
    \right) \neq \mathbb{0},
\end{equation}
\end{widetext}
by a simple reordering of indices (in the last step). With the same reasoning as above, we focus our attention to the right-hand side of the tensor product,
\begin{equation}
    \sum_{kn = 0}^{d-1} \ket{k}\!\!\bra{n}
    \left(
    \sum_{r = 0}^{R} K^{(r)}_{ki} K^{(r)\dagger}_{jn} - \sum_{r = 0}^{R'}K'^{(r)}_{ki} K'^{(r)\dagger}_{jn}
    \right) \neq \mathbb{0}_D,
\end{equation}
since the left-hand side is again trivially satisfying the non-zero condition. In this form, it is clear that the condition is satisfied if for a single set of indices, $i,j,k,l$, we have that 
\begin{equation}
    \sum_{r = 0}^{R} K^{(r)}_{ki} K^{(r)\dagger}_{jn} - \sum_{r = 0}^{R'}K'^{(r)}_{ki} K'^{(r)\dagger}_{jn}
    \neq 0.
\end{equation}

\section{Gauge Freedoms Interpreted in the Kraus Representation}\label{app:gauge_freedoms}
To understand how the action of any spoofer (either Type-I or Type-II), $\mathcal{E}'$, affects the channel in more physical terms, consider the Kraus representation of $\mathcal{E}'$. To change representation, consider the eigenvalue decomposition of $J_{\mathcal{E}'}$,
\begin{equation}
    J_{\mathcal{E}'} = \sum_{j} J_j' \ket{e_j}\!\!\bra{e_j}.
\end{equation}
Using the spectral theorem, we may express the eigenvectors, $\ket{e_j}$, as 
\begin{equation}
    \ket{e_j} = (K_j' \otimes \mathbb{I}) \ket{\Omega},
\end{equation}
where $\ket{\Omega} = \sum_{k}\ket{kk}$ is the un-normalised maximally entangled state. In order to extract the Kraus operator from our eigenvector, we can apply a computational basis decomposition,
\begin{equation}
    \ket{e_j} = \sum_{il}c_{il}^j \ket{il},
\end{equation}
from which we see that
\begin{equation}
    \braket{il}{e_j} = c_{il}^j = \bra{il}(K_j \otimes \mathbb{I}) \ket{\phi} = \braket{i}{K_j|l}.
\end{equation}

\section{Rank-reduction for 1-qubit Pauli channels}\label{app:1qubit}
As an example of the algorithm presented in Sec.~\ref{sec:analytic_pauli_lowering}, consider its action on an arbitrary 1-qubit Pauli channel $J_{\mathcal{E}} \mapsto J_{\mathcal{E}'}$ as
\begin{equation}
\begin{split}
     J_{\mathcal{E}}=&
    \begin{pmatrix}
        \alpha_0 + \alpha_3 & 0 & 0 & \alpha_0 - \alpha_3 \\
        0 & \alpha_1 + \alpha_2 & \alpha_1 - \alpha_2 & 0 \\
        0 & \alpha_1 - \alpha_2 & \alpha_1 + \alpha_2 & 0 \\
        \alpha_0 - \alpha_3 & 0 & 0 & \alpha_0 + \alpha_3
    \end{pmatrix} \\
    & \mapsto 
    J_{\mathcal{E}'} = 
    \begin{pmatrix}
        \alpha_0' & 0 & 0 & \alpha_0' \\
        0 & \alpha_1' & \alpha_1' & 0 \\
        0 & \alpha_1' & \alpha_1' & 0 \\
        \alpha_0' & 0 & 0 & \alpha_0'
    \end{pmatrix},
\end{split}
\end{equation}
where $\alpha_0' = \alpha_0 + \alpha_3$ and $\alpha_1' = \alpha_1 + \alpha_2$.
We see that the new channel $\mathcal{E}'$ is in the same equivalence class as $\mathcal{E}$, but with Kraus rank equal to 2 instead of 4.

\end{document}

%% file: figures/channel_representations.tex
\begin{tikzpicture}[>=stealth, auto, line width = 1.5pt]
    \coordinate (Kraus) at (-2, 0);
    \coordinate (Natural) at (6, 0);
    \coordinate (Choi) at (2, {2*sqrt(3)});
    
    \node[rectangle, minimum width=1.5cm, minimum height=1cm] (KrausNode) at (Kraus) {Kraus (Operator Sum) \(\{K_j\}\)};
    \node[rectangle, minimum width=1.5cm, minimum height=1cm] (NaturalNode) at (Natural) {Natural (Superoperator) \(S_{\mathcal{E}}\)};
    \node[rectangle, minimum width=1.5cm, minimum height=1cm] (ChoiNode) at (Choi) {Choi (Dynamical) \(J_{\mathcal{E}}\)};
    
    \draw[->] ([yshift=4pt]KrausNode.east) -- node[midway, above]{$S = \sum_j K_j \otimes K_j^{*}$} ([yshift=4pt]NaturalNode.west);
    
    \draw[->] ([xshift=-20pt]ChoiNode.south) -- node[midway, right]{\textit{S.T}} ([xshift=-20pt]KrausNode.north east);
    \draw[->] ([xshift=-30pt]KrausNode.north east) -- node[midway, left]{\(\sum_k \text{vec}(K_j) \text{vec}(K_j)^{\dagger}\)} ([xshift=-30pt]ChoiNode.south);
    
    \draw[->] ([xshift=30pt]NaturalNode.north west) -- node[midway, right]{$R$} ([xshift=30pt]ChoiNode.south);
    \draw[->] ([xshift=20pt]ChoiNode.south) -- node[midway, left]{$R$} ([xshift=20pt]NaturalNode.north west);
\end{tikzpicture}

%% file: figures/spoofing_types.tex
    \centering
        \begin{tikzpicture}[scale=0.9, every node/.style={scale=0.9}]
        \node at (-0.5,0) {(a)};
        \node at (0,0) {\(\rho\)};
        \draw[black, thick] (0.3,0) -- (0.8,0);
        \node[draw, minimum width=1.5cm, minimum height=1.2cm] at (1.55,0) {\(\mathcal{E}\)};
        \node[draw, minimum size=1cm] at (3.5,0) {}; 
        \draw[black, thick] (2.3,0) -- (3.0,0); 
        \draw[black, thick] (3.15, -0.15) arc[start angle=180, end angle=0, radius=0.35];
        \draw[black, thick, ->] (3.40, -0.15) -- (3.80, 0.3);
        \node at (3.40, -0.8) {\(\mathcal{M}\)};
    
        \draw[black, thick, ->] (4.75,0) -- (7,0) node[midway, above] {\textbf{Type I Spoofing}};

        \node at (8,0) {\(\rho\)};
        \draw[black, thick] (8.3,0) -- (9,0);
        \node[draw, minimum width=1.5cm, minimum height=1.2cm] at (9.75,0) {\(\mathcal{E}\)};
        \draw[black, thick] (10.5,0) -- (10.75,0);
        \node[draw, minimum width=1.5cm, minimum height=1.2cm] at (11.5,0) {\(\Phi\)};
        \draw[black, thick] (12.25,0) -- (13.6,0);
        \node[draw, dashed, fit={(8.8,-0.8) (12.45,0.8)}] {};

         \draw[black, thick] (13.6,0) -- (13.7,0); 
        \node[draw, minimum size=1cm] at (14.2,0) {}; 
        \draw[black, thick] (13.85, -0.15) arc[start angle=180, end angle=0, radius=0.35];
        \draw[black, thick, ->] (14.15, -0.15) -- (14.40, 0.3);
        \node at (14.1, -0.8) {\(\mathcal{M}\)};

        \node at (10.1,-1.2) {\(\mathcal{E'} = \Phi \circ \mathcal{E}\)};



      
        \node at (-0.5,-2.5) {(b)};
        \node at (0,-2.5) {\(\rho\)};
        \draw[black, thick] (0.3,-2.5) -- (0.8,-2.5);
        \node[draw, minimum width=1.5cm, minimum height=1.2cm] at (1.55,-2.5) {\(\mathcal{E}\)};

        \node[draw, minimum size=1cm] at (3.5,-2.5) {}; 
        \draw[black, thick] (2.3,-2.5) -- (3.0,-2.5); 
        \draw[black, thick] (3.15, -2.65) arc[start angle=180, end angle=0, radius=0.35];
        \draw[black, thick, ->] (3.40, -2.65) -- (3.80, -2.2);
        \node at (3.40, -3.3) {\(\mathcal{M}\)};
    
        \draw[black, thick, ->] (4.75,-2.5) -- (7,-2.5) node[midway, above] {\textbf{Type II Spoofing}};

        \node at (8,-2.5) {\(\rho\)};
        \draw[black, thick] (8.3,-2.5) -- (9,-2.5);
        \node[draw, minimum width=2cm, minimum height=1.2cm] at (10,-2.5) {\(\mathcal{E}'\)};
        \draw[black, thick] (11,-2.5) -- (12.9,-2.5);
    

        \draw[black, thick] (12.9,-2.5) -- (13.0,-2.5); 
        \node[draw, minimum size=1cm] at (13.5,-2.5) {}; 
        \draw[black, thick] (13.15, -2.65) arc[start angle=180, end angle=0, radius=0.35];
        \draw[black, thick, ->] (13.45, -2.65) -- (13.70, -2.2);
        \node at (13.4, -3.3) {\(\mathcal{M}\)};

        \end{tikzpicture}

%% file: figures/labels_spoofer_geometry.tex
\begin{tikzpicture}

    \node[anchor=south west,inner sep=0] (image) at (0,0) {\includegraphics[width=10cm]{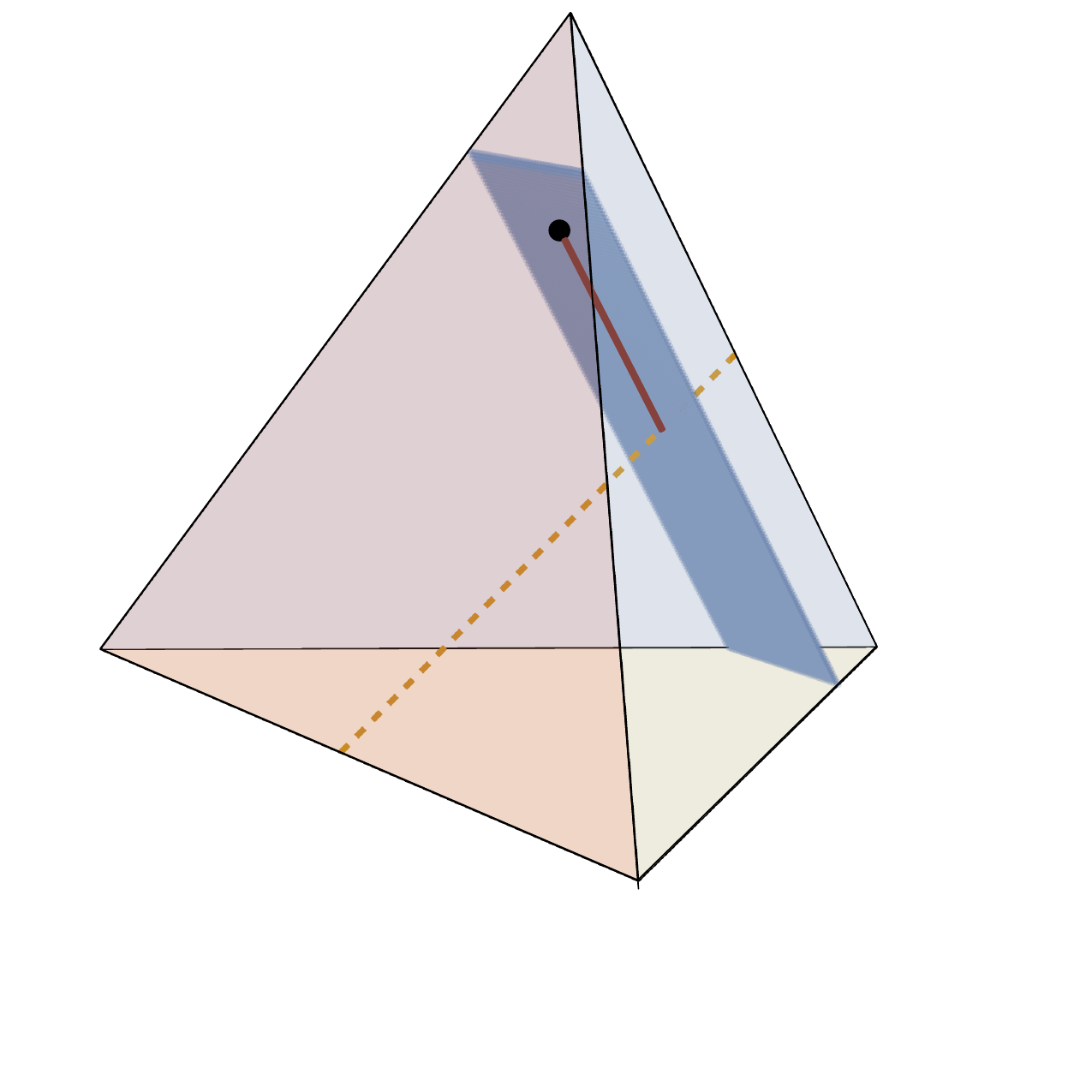}};
    \begin{scope}[shift={(image.south west)}, x={(image.south east)}, y={(image.north west)}]
        \node[anchor=south] at (0.52,1.0) {\(\mathbb{I}\)}; 
        \node[anchor=north west] at (0.03,0.43) {X};
        \node[anchor=north] at (0.85,0.43) {Z}; 
        \node[anchor=north] at (0.58,0.18) {Y}; 
    \end{scope}
    \begin{scope}[shift={(image.north east)}, xshift=-1.3cm, yshift=-1cm]
        \draw[rounded corners, fill=white, draw=black] (0,-0.9) rectangle (6.2,-3.3);
        
        \definecolor{linecolor}{RGB}{133, 65, 59}
        \draw[rounded corners, fill=linecolor] (0.2,-1) rectangle (0.6,-1.4);
        \node at (0.7,-1.2) [anchor=west] {Type-I Spoofing};
        
        \definecolor{planecolor}{RGB}{134, 156, 189}
        \draw[rounded corners, fill=planecolor] (0.2,-1.6) rectangle (0.6,-2);
        \node at (0.7,-1.8) [anchor=west] {Type-II Spoofing};

        \draw[rounded corners, fill=black] (0.2,-2.2) rectangle (0.6,-2.6);
        \node at (0.7,-3.0)
        [anchor=west] {Invariant Type-I Spoofing Line};

        \definecolor{dashedcolor}{RGB}{201, 135, 45}
        \draw[rounded corners, fill=dashedcolor] (0.2,-2.8) rectangle (0.6,-3.2);
        \node at (0.7,-2.4) [anchor=west] {Original Channel};
        
    \end{scope}
\end{tikzpicture}

%% file: TikZ/eigenvalues_convergence.tex
\begin{tikzpicture}[scale=0.854, every node/.style={scale=0.854}]
    \node (a) at (-8.2,3.2) {(a)};
    \node (a) at (0,0) {    \includegraphics[width=1\linewidth]{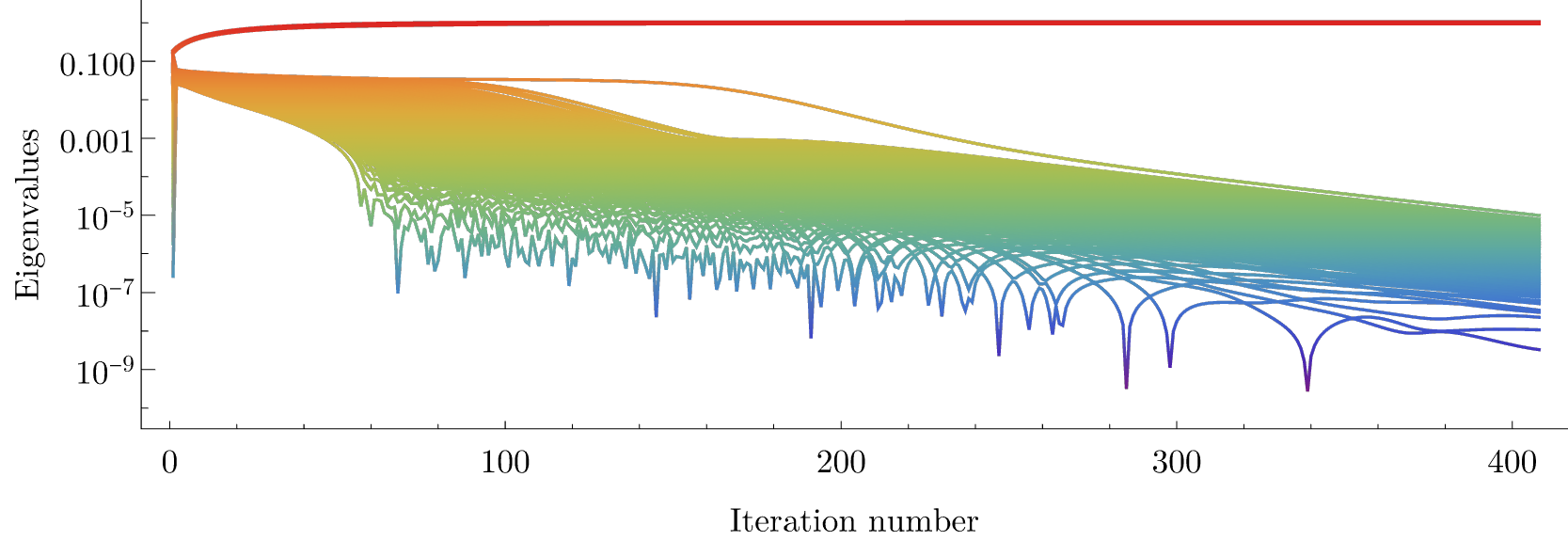}};
    \node (a) at (9.5,2.85) {$\{1,...,20\}$};
    \node (a) at (9.0,0.65) {21};
    \node (a) at (9,0,0.) {\vdots};
    \node (a) at (9.0,-0.9) {400};

    \node (a) at (-8.2,-3.3) {(b)};
    \node (a) at (0,-6.6) {    \includegraphics[width=1\linewidth]{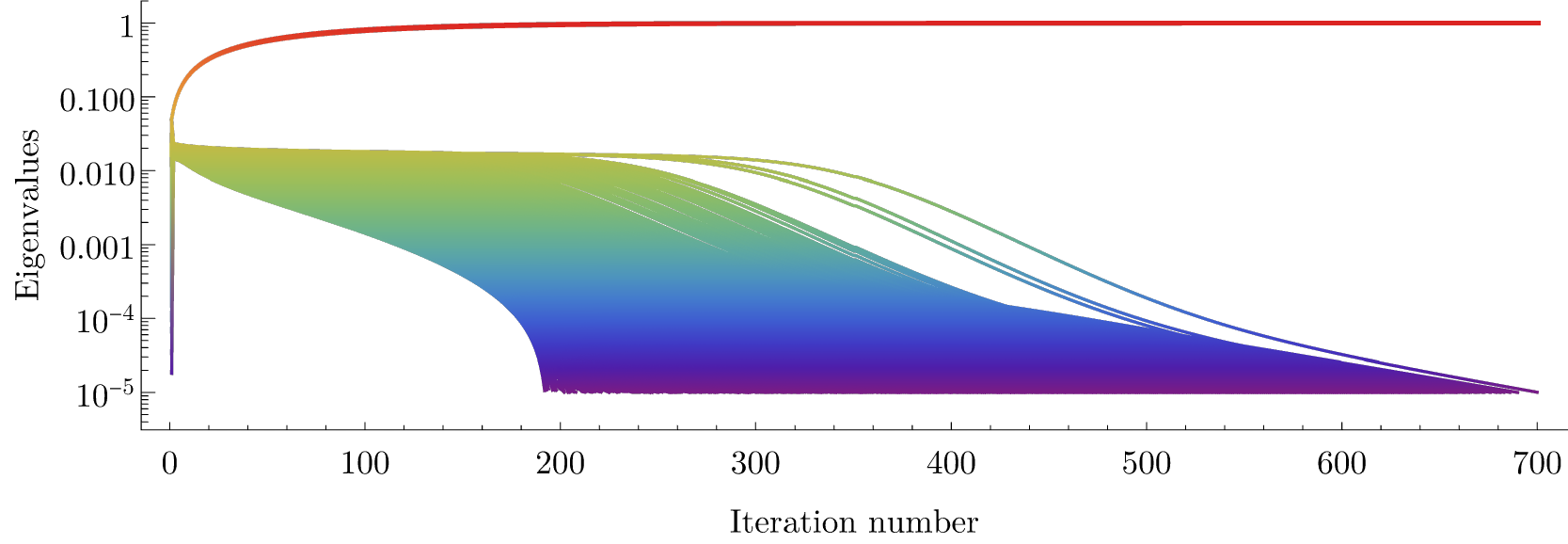}};
    \node (a) at (9.5,-3.77) {$\{1,...,50\}$};
    \node (a) at (9.7,-7.95) {$\{51,...,2500\}$};
\end{tikzpicture}

%% file: TikZ/non-zero_pattern_2q.tex
\begin{tikzpicture}[scale=0.885, every node/.style={scale=0.885}]
    \node (a) at (0,0) {\includegraphics[width=0.48\textwidth]{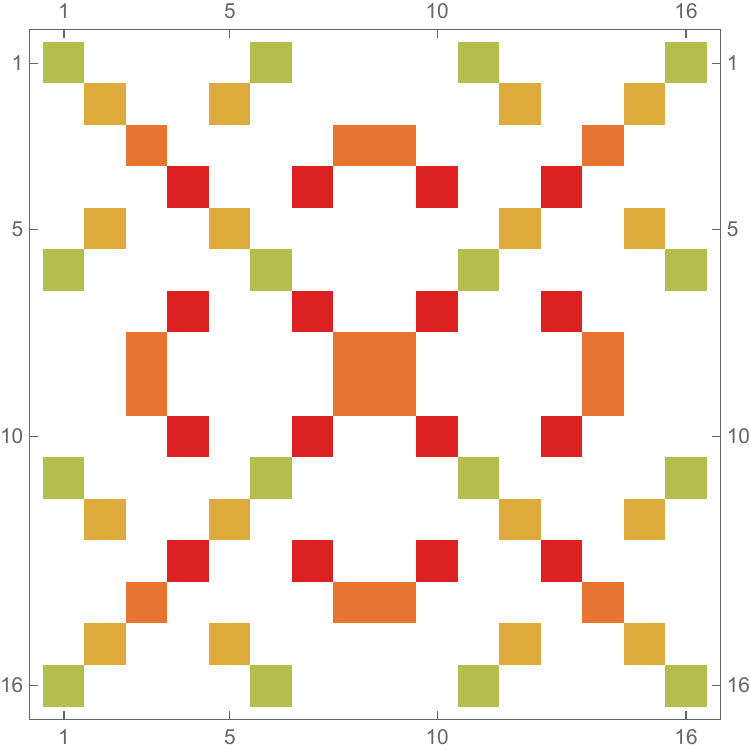}};
\end{tikzpicture}

%% file: TikZ/non-zero_pattern_3q.tex
\begin{tikzpicture}[scale=0.885, every node/.style={scale=0.885}]
    \node (a) at (0,0) {\includegraphics[width=0.48\textwidth]{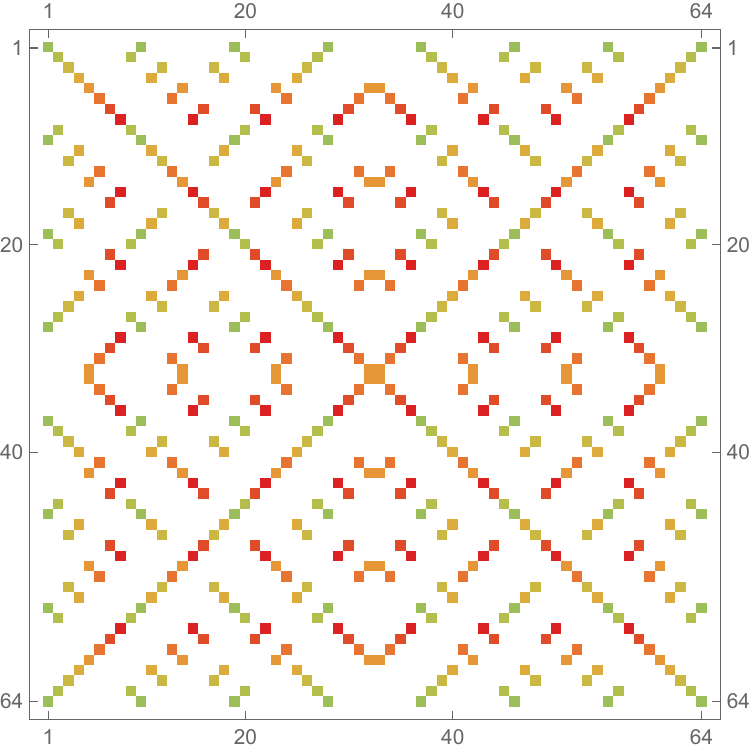}};
\end{tikzpicture}